\documentclass[12pt]{article}

\usepackage{graphicx}
\usepackage{amsmath}
\usepackage{amssymb}
\usepackage{multirow}

\allowdisplaybreaks
\def\eq#1{(\ref{#1})}
\def\[#1\]{\begin{align}#1\end{align}}

\begin{document}

\begin{titlepage}
\title{
\hfill\parbox{4cm}{ \normalsize YITP-16-2}\\ 
\vspace{1cm} 
Emergent classical geometries on boundaries \\ of randomly connected tensor networks}
\author{
Hua Chen$^a$\footnote{hua.chen@yukawa.kyoto-u.ac.jp}, 
Naoki Sasakura$^a$\footnote{sasakura@yukawa.kyoto-u.ac.jp}, 
Yuki Sato$^{b}$\footnote{Yuki.S@chula.ac.th}
\\
$^{a}${\small{\it Yukawa Institute for Theoretical Physics, Kyoto University,}}
\\ {\small{\it  Kitashirakawa, Sakyo-ku, Kyoto 606-8502, Japan,}}
\\
$^{b}${\small{\it Department of Physics, Faculty of Science, Chulalongkorn University}}
\\ {\small{\it Thanon Phayathai, Pathumwan, Bangkok 10330, Thailand.}}\\
}

\date{\today}
\maketitle
\thispagestyle{empty}
\begin{abstract}
\normalsize
It is shown that classical spaces with geometries emerge on boundaries
of randomly connected tensor networks  
with appropriately chosen tensors in the thermodynamic limit.
With variation of the tensors,
the dimensions of the spaces can be freely chosen, 
and the geometries, which are curved in general, can be varied.
We give the explicit solvable examples of emergent flat tori in arbitrary dimensions,
and the correspondence from the tensors to the geometries for general curved cases.
The perturbative dynamics in the emergent space is shown to be 
described by an effective action 
which is invariant under the spatial diffeomorphism due to the underlying orthogonal group 
symmetry of the randomly connected tensor network.
It is also shown that there are various phase transitions among spaces,
including extended and point-like ones, under continuous change of the tensors.
\end{abstract}
\end{titlepage}

\section{Introduction}
\label{sec:intro}
The construction of quantum gravity is one of the most fundamental problems in theoretical physics.
Efforts made for this problem so far 
suggest that the classical notion of 
spacetime described by general relativity 
should somehow be replaced  with a new one suitable 
for quantization. 
The classical notion, namely the continuous entity of spacetime
with geometries, 
should appear as an infrared effective description of the new picture.
The main subject of this paper is to explicitly show that the classical notion of spaces with geometries
appears as an emergent phenomenon 
from the dynamics of the randomly connected tensor network.  
Here, the randomly connected tensor network is defined as random summation
of tensor networks: 
tensors are treated as controllable external variables, while connections of tensors 
are randomly summed over. 
We will show that spaces with geometries appear on boundaries of randomly connected tensor networks,
if tensors are appropriately chosen. 
By varying the tensors, one can freely choose the dimensions of the spaces,
and can also vary the geometries.
We will explicitly give the  correspondence from the tensors to the geometries for 
general curved cases.

The background motivation for this work comes from the fact that the randomly connected tensor network is tightly 
related to a tensor model in the Hamilton formalism.
Tensor models \cite{Ambjorn:1990ge,Sasakura:1990fs,Godfrey:1990dt}
have originally been introduced to describe $D>2$ simplicial quantum gravity
as extensions of the matrix models which successfully describe the $D=2$ simplicial quantum 
gravity.\footnote{Interestingly, a matrix-model-like approach to 
$D=3$ simplicial quantum gravity has recently appeared \cite{Fukuma:2015xja,Fukuma:2015haa}.} 
Though the original tensor models suffer from some difficulties,
the coloured tensor models \cite{Gurau:2009tw}, which appeared later with
improvements, 
have extensively been analyzed \cite{Gurau:2011xp}. However, 
the analyses so far have shown that the dominant configurations do not generate realistic
structures comparable to our actual spacetime: branched polymers, for example, dominate 
\cite{Gurau:2011xp,Bonzom:2011zz,Gurau:2013cbh} in the simplest settings.
On the other hand, the numerical analyses of $D>2$ simplicial quantum gravity have shown that
the Lorentzian models called causal dynamical triangulations (CDT) successfully generate de-Sitter like spacetimes
like our actual universe \cite{Ambjorn:2004qm}, while the Euclidean ones do 
not.\footnote{When coupling many U$(1)$-fields, 
the authors in \cite{Horata:2000eg} found a promise of 
a phase transition higher than first order, 
which, however, is in conflict with the result in \cite{Ambjorn:1999ix}.}
This success of CDT 
motivated one of the present authors to formulate a tensor model in the Hamilton 
formalism \cite{Sasakura:2011sq,Sasakura:2012fb}, 
which we call canonical tensor model (CTM). 
CTM has been shown to have various interesting properties.
It is unique  under reasonable assumptions \cite{Sasakura:2012fb},
 and has tight connections with general relativity \cite{Sasakura:2014gia,Sasakura:2015pxa}.
In addition, it has connections to the randomly connected tensor network \cite{Dorogovtsev:2008zz,Sasakura:2014yoa}:
the Hamiltonian of CTM generates
a sort of a renormalization group flow of the randomly connected tensor 
network \cite{Sasakura:2014zwa,Sasakura:2015xxa}. 
The randomly connected tensor network is also useful in constructing the exact physical states 
of the quantized version of CTM \cite{Narain:2014cya,Sasakura:2013wza}.

So far, a number of network-based models for emergent space or spacetime 
have been proposed, \textit{e.g.}  
spin networks \cite{Penrose:1971}, 
loop quantum gravity (see \textit{e.g.} \cite{Rovelli:2014ssa}), 
causal sets \cite{Sorkin:1987, Sorkin:1999}, energetic causal sets \cite{Cortes:2013uka, Cortes:2013pba}, 
quantum graphity \cite{Konopka:2006hu, Konopka:2008hp, Hamma:2009xb}, 
information-bits model by Trugenberger \cite{Trugenberger:2015xma, Trugenberger:2015maa}, 
Wolfram's evolving networks \cite{Wolfram:2002}, D'Ariano-Tosini causal networks \cite{D'Ariano:2010dr}, 
structurally dynamic cellular networks (see \textit{e.g.} \cite{Requardt:2015ila}), 
lumpy networks (see \textit{e.g.} \cite{Requardt:2003qy}), 
complex quantum network manifold \cite{Bianconi:2015hfa, Bianconi:2015wla} 
and network geometry with flavor \cite{Bianconi:2015qpp}. 
All the above constructions capture aspects of geometry from the bulk of networks. 
In contrast, in our model not the bulk but boundaries of networks provide the information of geometry.  

This paper is organized as follows. In Section~\ref{sec:RCTN}, we define the randomly connected tensor network, and 
explain our method to analyze it, which has been developed in \cite{Sasakura:2014yoa,Sasakura:2014zwa,Sasakura:2015xxa}. 
In Section~\ref{sec:space}, we explain some difficulties in regarding the randomly connected tensor network itself as a space,
and argue that its boundary is more qualified.
In Section~\ref{sec:flat}, we explicitly give such tensors that generate flat spaces of arbitrary dimensions
on the boundaries of the randomly connected tensor networks.
In Section~\ref{sec:general}, we consider variation of the tensors 
from those of the flat spaces
to generate curved spaces. We explicitly give the correspondence from the tensors 
to the geometries of the spaces.
In Section~\ref{sec:phasetransition}, we study phase transitions in which 
the flat spaces break down to point-like spaces and others.
The final section is devoted to the summary and discussions.
In Appendix~\ref{sec:covariance}, we describe
a covariant structure under the spatial diffeomorphism
in the continuum description of the randomly connected tensor network,
which is used in Section~\ref{sec:general}.

\section{Randomly connected tensor network}
\label{sec:RCTN}
\begin{figure}
\begin{center}
\includegraphics[width=3cm]{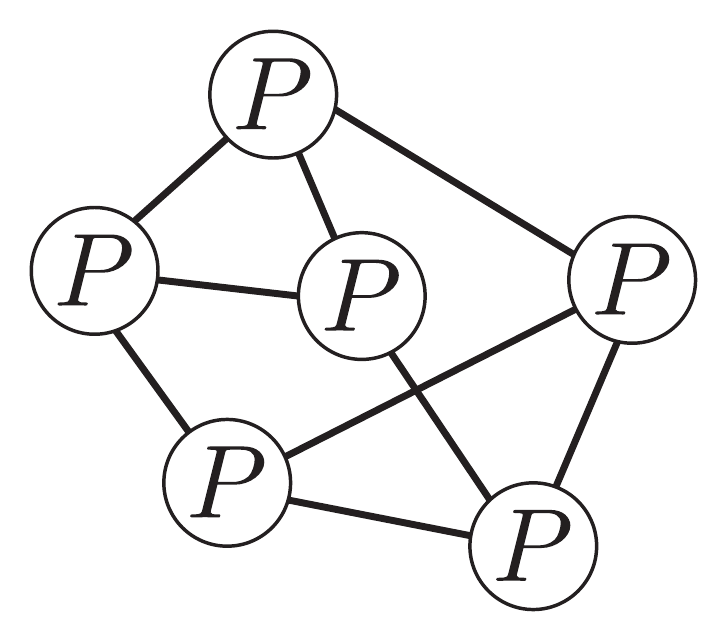}
\caption{An example of a tensor network with $n=6$.
\label{fig:networkP}}
\end{center} 
\end{figure}

We consider a real symmetric rank-three tensor, $P_{abc}$,  
where $a,b,c=1,2, \cdots, N$. As a tensor product of $P_{abc}$'s 
we introduce the following rank-$3n$ tensor:
\[
W_{a_1 a_2 \cdots  a_{3n}} (P) 
= P_{a_1 a_2 a_3} P_{a_4 a_5 a_6} \cdots P_{a_{3n-2} a_{3n-1} a_{3n}},
\label{eq:w}
\] 
where $n$ is even integer. 
Contracting all indices in (\ref{eq:w}) in pairs by $\frac{3}{2}n$ delta functions $\delta_{a_i a_j}$, 
we construct O($N$)-invariant quantities.
We denote such O($N$)-invariant quantities by $W_{g}(P)$ 
where $g$ specifies one way of contracting all indices in pairs. 
In fact, one can consider $g$ as a regular network consisting of $n$ three-vertices. 
Namely, we introduce $n$ trivalent vertices and assign $P_{abc}$ to each vertex in such a way that 
three lines emerging from a single trivalent vertex carry the indices, $a$, $b$ and $c$, respectively, 
and connect them by lines if the indices are contracted. 
In this way, one can construct a regular network $g$ allowing self-contractions.
The self-contraction means that two of the three lines originating from a single vertex are connected by a line. 
See Fig.\ref{fig:networkP} for an example of the networks. 
By this understanding, 
we define the partition function as a sum over all networks $g$: 
\[
Z_n (P) 
= \sum_{g} \frac{1}{|\text{Aut} (g)|} W_{g} (P),
 \label{eq:znp01} 
\] 
where $|\text{Aut} (g)|$ is the order of the automorphism group of $g$. 
We call the model defined by the partition function (\ref{eq:znp01}) 
\textit{randomly connected tensor networks}, 
and the randomness comes from the sum over all networks 
with the tensor assigned to each vertex. 
In fact one can choose $P_{abc}$ 
so that the partition function (\ref{eq:znp01}) describes a statistical system
on random networks \cite{Dorogovtsev:2008zz,  Sasakura:2014yoa,Sasakura:2014zwa},
such as the Ising model on random networks. 
 
For convenience, we rewrite the partition function (\ref{eq:znp01}) 
in terms of the real integral 
\cite{Sasakura:2014zwa,Sasakura:2014yoa,Sasakura:2015xxa}: 
\[
Z_n(P)=\frac{1}{n!}\int_{-\infty}^\infty
\frac{\text{d}^N \varphi }{ (2\pi)^{N/2} } 
\left( \frac{1}{3!} P \varphi^3 \right)^n e^{-\frac{1}{2} \varphi^2},
\label{eq:ZofRCTN}
\]
where $\varphi_a\ (a=1,2,\ldots,N)$ are real integration 
variables, and we have used the shorthand notations given by
\[
P \varphi^3 := P_{abc} \varphi_a \varphi_b \varphi_c, \ \ \
 \varphi^2 := \varphi_a \varphi_a, \ \ \ 
 \text{d}^N \varphi := \prod^N_{a=1} \text{d} \varphi_a. 
\label{eq:short1}
\]
Here, 
repeated indices above are assumed to be summed over, and hereafter
 we will use this convention for the sum. 
Applying Wick's theorem, one can confirm that the partition functions, (\ref{eq:znp01}) and (\ref{eq:ZofRCTN}), 
coincide. The partition function \eq{eq:ZofRCTN} is invariant under the O($N$) transformation of $P$,
\[
P'_{abc}=L_{a}{}^{a'}L_{b}{}^{b'}L_{c}{}^{c'}P_{a'b'c'},\ \ L\in \text{O}(N).
\label{eq:orthogonal}
\]

Let us consider the thermodynamic limit of the randomly connected tensor network, 
in which the size of the networks grows to infinity, $n\rightarrow \infty$.
It can be shown that the partition function \eqref{eq:ZofRCTN} in this limit can be exactly computed by a mean field
method \cite{Sasakura:2014zwa,Sasakura:2014yoa,Dorogovtsev:2008zz}. 
To see this, 
we implement the rescaling, 
$
\varphi_a \to \varphi'_a = \sqrt{2n} \ \phi_a,
$
and correspondingly the partition function (\ref{eq:ZofRCTN}) becomes
\[
Z_{n}(P) 
 = C_n 
  \int^{\infty}_{-\infty}
 \text{d}^N \phi 
 \ e^{-n f(P, \phi)}, 
 \label{eq:ZofRCTN2}
\]
where
\[
C_n 
=\frac{(2n)^\frac{3n}{2}}{n! \left( 3! \right)^n} \left( \frac{n}{\pi} \right)^\frac{N}{2},
\]
\[
f(P, \phi) = \phi^2 - \frac{1}{2} \log 
\left[  
A(\phi)^2
\right]
, 
\ \ \ \text{with} \ \ \ 
A(\phi) 
= 
P \phi^3
. 
\label{eq:freeenergywithA}
\]
From the expression \eq{eq:ZofRCTN2}, in the thermodynamic limit $n\rightarrow\infty$,
the steepest descent method\footnote{In real valued cases like here, the method is also called the Laplace method.}
can be applied, and the partition function is determined by the neighborhood of  the minimum of
$
f(P,\phi) 
$
as a function of $\phi$. 
Thus the free energy per vertex in the thermodynamic limit can be defined as
\[
f(P) 
:= 
-\lim_{n\rightarrow \infty} \frac{1}{n} \ln\left( \frac{Z_n(P)}{C_{n}}\right)
=
\min_\phi f(P,\phi) 
= f(P,\phi_{min}).
\label{eq:freeenergy}
\]
In the expression above 
we have removed the numerical factor $C_{n}$ from the definition of the free energy,
since this simply gives a $P$-independent shift of the free energy. 
The minimum $\phi=\phi_{min}$ is one of the solutions $\phi = \bar \phi$ to the stationary condition,
\[
\left.\frac{\partial f(P,\phi)}{\partial \phi_a} \right|_{\phi=\bar \phi}
=2\bar \phi_a-\frac{3 (P\bar \phi^2)_a}{A(\bar \phi )}  =0,
\label{eq:stationary}
\]  
where we have used the following shorthand notation:
\[
(P\phi^2)_a := P_{abc} \phi_b \phi_c.
\]
Multiplying \eq{eq:stationary} by $\bar \phi$, 
one obtains
\[
\bar \phi^2=\frac{3}{2}.
\label{eq:phi2}
\]
In simplifying expressions, \eq{eq:stationary} is useful in the form,
\[
(P\bar \phi^2)_a=\frac{2}{3} A(\bar \phi ) \bar \phi_a.
\label{eq:simple}
\]
Note that $\bar \phi$ and $\phi_{min}$ are generally dependent on $P$, but we suppress the argument for notational simplicity. 

In general, in the vicinities of first order phase transition surfaces, 
not only the absolute minimum of the free energy
but also stable local minima are physically relevant. 
The local stability of each stationary point can be checked by evaluating the
positivity of the Hessian, {\it i.e.} the matrix of the second order derivatives of $f(P,\phi)$ with respect to $\phi$ evaluated at $\phi=\bar \phi$.
Using \eqref{eq:simple} to simplify the expression, the Hessian is obtained as 
\[
f^{(2)}_{ab} := \left. \frac{\partial^2 f(P,\phi)}{\partial \phi_a \partial \phi_b} \right|_{\phi=\bar \phi} 
=2 \left( \delta_{ab} +2 \bar \phi_a \bar \phi_b -\frac{3 (P\bar \phi)_{ab}}{ A(\bar \phi )} \right).
\label{eq:f2ab}
\]
One of its eigenvectors is $\bar \phi$ with eigenvalue $4$:
\[
f^{(2)}_{ab} \bar \phi_b=2 \left( \bar \phi_a +2 \bar \phi^2 \bar \phi_a 
-\frac{3 (P\bar \phi^2)_a }{ A(\bar \phi ) } \right) 
=4\bar \phi_a,
\]
where we have used \eqref{eq:phi2}, \eqref{eq:simple}, and \eqref{eq:f2ab}.

\section{Boundaries of random networks as spaces}
\label{sec:space}
It is not an easy question how random networks can be regarded as spaces.
The most naive manner of identification would be to regard vertices as 
points forming a space and connecting lines as representations of local structures of 
their neighbourhoods. 
In fact, this is the common perspective used in lattice theories.  
However, this lattice-like way of viewing random networks as spaces seems to have at 
least the following difficulties:    
\begin{itemize}
\item{\it Difference of structures between random networks and our space}\,:  
Random networks do not seem to have similar structures 
as our actual space, which is smooth, respects locality, and has three dimensions. 
In fact, it is known that, in the thermodynamic limit,  
random networks with a fixed degree of vertices
effectively approach the Bethe lattice, which is tree-like \cite{Dorogovtsev:2008zz}.
This is far from the actual structure of our space.
\item {\it Finite correlation lengths}\,: 
The Ising model on random networks with a finite degree of vertices
has a finite correlation length even on the second order 
phase transition point in the thermodynamic limit \cite{Dorogovtsev:2008zz}.
Since this seems to be an outcome of the tree-like structure of the Bethe lattice in the thermodynamic limit,
finiteness of correlation lengths will generally hold in the other statistical systems
on random networks. 
This means that it is not possible to get any modes which propagate infinitely far.
Therefore, it is not plausible to obtain physically sensible theories for the actual space
from such a framework.
\item{\it Difficulty in labeling positions in random networks}\footnote{NS would 
like to thank N.~Kawamoto for stressing this difficulty in our framework 
at a workshop, ``Discrete approaches to dynamics of fields and spacetime",
held at Okayama, Japan,  in September, 2015. }:  
In our framework, each vertex is just a representation of a tensor $P$.
Therefore, the vertices are basically all the same and 
cannot be distinguished from each other in a well-defined way
in the random sum over networks.  
This difficulty in labelling the vertices would be an obstacle in obtaining 
any classical geometries.
\end{itemize}

\begin{figure}
\begin{center}
\includegraphics[scale=1]{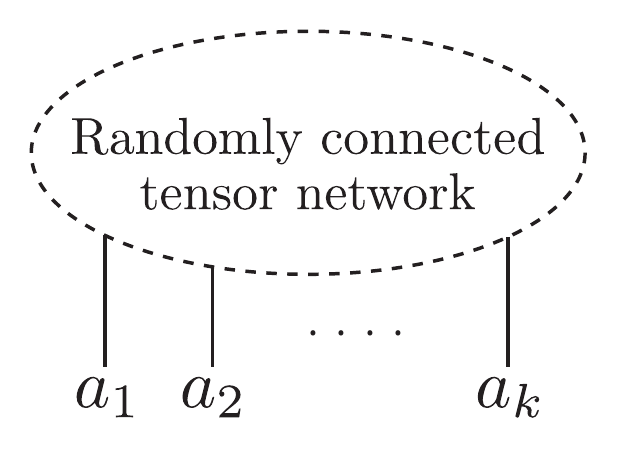}
\caption{The graphical representation of the correlation function
$\langle
 \varphi_{a_1}  \varphi_{a_2} \cdots  \varphi_{a_k} 
\rangle$. }
\label{fig:external}
\end{center}
\end{figure}

In view of these difficulties, the naive picture to identify randomly connected tensor networks
as spaces may not be a right way. 
Especially, the last difficulty above suggests that we should more 
seriously consider what are the appropriate quantities to observe 
in randomly connected tensor networks. 
The natural physically relevant quantities 
of the system defined by the partition function \eqref{eq:ZofRCTN} are 
the correlation functions of $\varphi$:
\[
\left\langle
\varphi_{a_1} \varphi_{a_2} \cdots \varphi_{a_k} 
\right\rangle
= 
\frac{1}{Z_n} 
\int_{-\infty}^\infty 
\frac{\text{d}^N \varphi}{(2 \pi)^{N/2}} 
\,
\varphi_{a_1} \varphi_{a_2} \cdots \varphi_{a_k}  
\,
\frac{1}{n!} \left( \frac{1}{3!} P \varphi^3 \right)^n e^{-\frac{1}{2}  \varphi^2}.
\label{eq:correlation}
\]
Applying Wick's theorem as before,  
the correlation function above corresponds 
to the summation 
over tensor networks 
with external lines having fixed indices (See Fig.\ref{fig:external}).
In other words,  
this is the random summation over tensor networks with fixed configurations
on their boundaries. 
So, we are pursuing the possibility of regarding boundaries of random networks
as spaces rather than networks themselves, or, in other words,
the possibility that ``points" labelled by the index set on the boundaries
form a space. Note that, due to the random summation over networks, the local structure of neighborhoods on such boundaries 
may be significantly different from that of networks themselves. This difference may solve the geometrical difficulties
mentioned in the first and second items of the list above. 
In the rest of this paper, 
we will show that, by appropriate choices of the tensor $P$, the correlation
functions behave as if $\varphi$ is a field in spaces with regular properties,
\textit{i.e.} 
the spaces are smooth, respect locality\footnote{However, as we will show later, 
there exists one non-local condition that a single mode of $\varphi$, 
the zero mode in a rough sense, becomes super-massive.
All the other modes are not conditioned. }  
and have certain dimensions.

To compute the correlation function \eqref{eq:correlation} in the 
thermodynamic limit, we investigate the partition function. 
To begin, we implement the rescaling, $\varphi \to \varphi' = \sqrt{2n} \phi$, 
as before, and perturb $\phi$ around one of local minima $\bar{\phi}$ defined by (\ref{eq:stationary})  as
\[
\phi_a = 
\bar{\phi}_a 
+ \frac{v_a}{\sqrt{n}}. 
\label{eq:expand}
\] 
Inserting (\ref{eq:expand}), the partition function (\ref{eq:ZofRCTN2}) for the large-$n$ 
can be expressed as
\[
C_n n^{- \frac{N}{2} } e^{-n f(P,\bar\phi)} \int_{-\infty}^\infty \text{d}^Nv \, 
e^{-\frac{1}{2}f^{(2)}_{ab} v_a v_b+\mathcal{O} \left(  \frac{v^3}{\sqrt{n}}
 \right) +\cdots }. 
\label{eq:partbyy}
\]

From \eq{eq:partbyy} one can see that, in the
thermodynamic limit, 
the system is described by a free theory of $v$. 
For instance, 
the connected part of the two point correlation function of $\varphi$ 
is given by
\[
\begin{split}
\left\langle \varphi_a \varphi_b \right\rangle _{\text{con.}} 
& = \left\langle \varphi_a  \varphi_b \right\rangle 
- \left\langle  \varphi_a \right\rangle \left\langle  \varphi_b \right\rangle \\  
&=2 \left\langle v_a v_b \right\rangle_{\text{con.}} \\
&=2 f^{(2)\, -1}_{ab} + \mathcal{O} \left( n^{-1} \right),
\end{split}
\label{eq:2pt}
\]
where $f^{(2)\, -1}$ is the inverse matrix of $f^{(2)}$.
Here we have also used that, from \eqref{eq:partbyy}, the one point function 
is sub-dominant, $\langle v_a \rangle \sim \mathcal{O}\left( n^{-1/2} \right).$
Higher order correlation functions behave similarly as a free theory in the leading order.

Thus, the challenge is whether we can find appropriate $P$'s
so that the correlation function \eqref{eq:2pt} behaves in the same manner 
as that in a regular space. This task contains the dynamical complexity that 
 the minimum solution $\bar \phi$ depends non-trivially on $P$. 
In Section~\ref{sec:flat}, we will give exactly solvable cases 
representing flat $D$-dimensional tori, and, in Section~\ref{sec:general}, we 
will treat spaces with general metrics.    
The dynamical complexity concerning the stability of the spaces
will be discussed in Section~\ref{sec:phasetransition}.  

\section{Flat spaces}
\label{sec:flat}
In this section, we explicitly give $P$'s  which realize arbitrary dimensional flat spaces.
The criterion for the emergence 
is whether the correlation function \eqref{eq:2pt} is similar to that
in a flat space. This is equivalent to showing that the eigenvalues and eigenvectors 
of $f^{(2)}$ are similar to those in a flat space.
We will first discuss the one-dimensional case, circles, and then consider
$D$-dimensional flat tori.

\subsection{Circles}
\label{subsec:circle}
  We parametrize $P_{abc}$ as 
  \begin{equation}
    \label{eq:configp}
    P_{abc} = P^L_{abc} +\xi  \left(P^L_{add} \delta_{bc} +P^L_{bdd}
      \delta_{ca} +P^L_{cdd} \delta_{ab}\right),
  \end{equation}
  where $\xi$ is a real parameter, and $P^L_{abc}$ is a local part of the tensor.
  The components of the local part are given by
  \begin{equation}
    \label{eq:P-local}
    P^L_{i,i,i} = 1, \quad P^L_{i,i+1,i+1} = P^L_{i,i,i+1} = \kappa,\ \ \ i=1,2,\ldots,N.
  \end{equation}
Here we use Latin $i,j,k$ on indices for components of a tensor, and 
the repetition of them does not imply summing over.
These indices are identified in modulo $N$, namely $N\sim 0$, to consider a circle composed of $N$ ``points''.  $\kappa$ is a real parameter.
  Then,
  \begin{eqnarray}
  \begin{split}
    \label{eq:P-specific}
    P_{iii} & = 1 +3\gamma_N, \\
    P_{ijj} & = 
                    \begin{cases}
                      \kappa +\gamma_N & i \sim j \\
                      \gamma_N & \text{otherwise}
                    \end{cases} , \\
    P_{ijk} & =  0,
    \end{split}
  \end{eqnarray}
  where  $i, j, k$ are always supposed to be different from each other, and $\gamma_N=\xi\sum_jP^L_{ijj}=\xi(1+2\kappa)$. The symbol $i\sim j$
  denotes a relationship of $i$ being $j$'s neighbour, which, in this
  case, means $i=j \pm1$ (modulo $N$).
  As can be seen in \eqref{eq:P-specific}, the term proportional to $\xi$ in 
  \eq{eq:configp} generates a non-local part of  $P$.
  
Because of the discrete translational symmetry in $P$,
a solution to the stationary condition  \eqref{eq:stationary} is given by
\[
\bar \phi^\pm_a=\pm\sqrt{\frac{3}{2 N}}.
\label{eq:flat1dim}
\]   
This solution is stable as a local minimum, when the spectra being obtained below are all positive.
On the other hand, whether this is an absolute minimum or not is a complex problem, 
and will be studied in Section~\ref{sec:phasetransition}.  

For the solution \eq{eq:flat1dim}, it is not hard to calculate that 
  \begin{equation}
    \label{eq:A-average}
    A(\bar \phi^\pm) = \pm N^{-1/2}\left(\frac{3}{2}\right)^{3/2}
    (1 +6\kappa +3\gamma),
  \end{equation}
where $\gamma=N\gamma_N$. 
  
  For the convenience of our future discussion, we introduce two symbols.
  $\tilde{\delta}_{ab}$ equals $1$ when $a \sim b$, and vanishes otherwise.
  $1_{ab}$ constantly equals to unit. Using these symbols, we get
  \begin{equation}
    \label{eq:PabcPhic}
    P_{abc}\bar \phi_c^\pm = \pm N^{-1/2}\left(\frac{3}{2}\right)^{1/2}
    \left[(1 +2\kappa +\gamma)\delta_{ab} +2\kappa\tilde{\delta}_{ab}
    +\frac{2\gamma}{N} 1_{ab}\right]. 
  \end{equation}
  Let us introduce
  \begin{equation}
    \label{eq:pqr}
    p = \frac{1 +2\kappa +\gamma}{1 +6\kappa +3\gamma} \,,\quad q = \frac{4\kappa}{1 +6\kappa
      +3\gamma} \,,\quad r = \frac{2\gamma}{1 +6\kappa +3\gamma}.
  \end{equation}
  They satisfy $p+q+r=1$. Then the Hessian \eqref{eq:f2ab}  can be computed as
  \begin{equation}
    \label{eq:free-energy}
    f^{(2)}_{ab} = (2 -4p) \delta_{ab} -2q \tilde{\delta}_{ab}
    +\frac{6-4r}{N} 1_{ab}.
  \end{equation}

\begin{figure}
 \begin{center}
 \includegraphics[scale=.7]{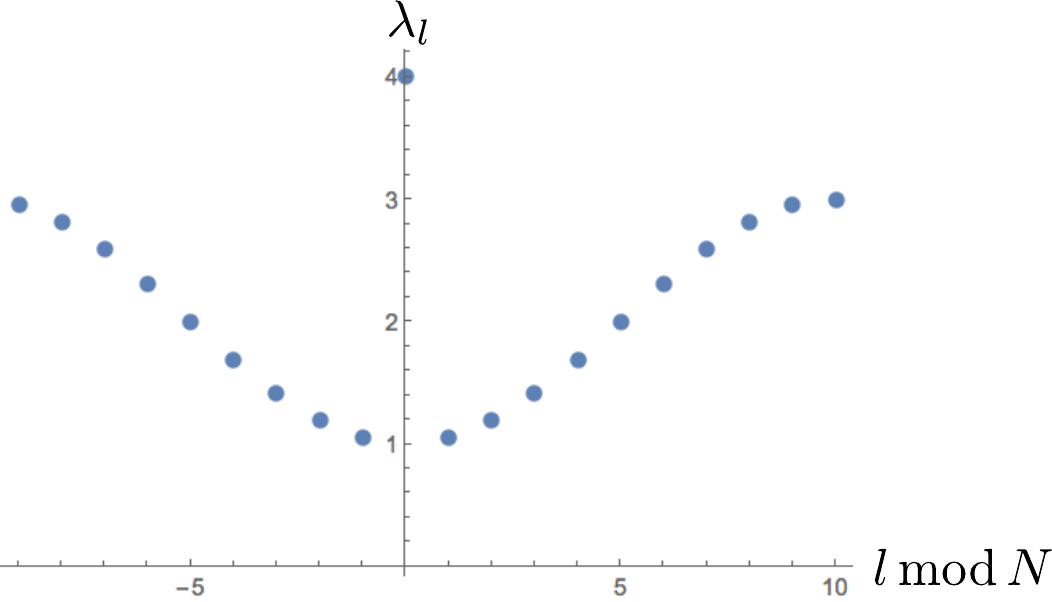}
 \caption{An example of the spectra \eq{eq:spectra} for $N=20,\, p=0,\,q=\frac{1}{4}$. }
 \label{fig:spectra}
 \end{center}
 \end{figure}

  Now, let us consider discrete analogues of plane waves,  $\psi^{(l)}_a=e^{i\frac{2\pi
      l}{N}a}$. We find that they are the eigenvectors of $f^{(2)}$: 
  \begin{equation}
    \label{eq:eigenvector}
    f^{(2)}_{ab}\psi^{(l)}_{b} = \left(2 -4p -4q\cos \frac{2\pi
          l}{N} \right) \psi^{(l)}_a \,,\quad 
    \text{for $l=1,2,\dots,N-1$},
  \end{equation}
  and $f^{(2)}_{ab}\psi^{(0)}_b=f^{(2)}_{ab}1_b=4\cdot1_a$. Thus, naming
  \begin{equation}
    \label{eq:spectra}
    \Lambda = \left\{ \left. \lambda_l = 2 -4p -4q \cos \frac{2\pi
          l}{N} \,\right|\, l = 1, 2, \dots, N-1\right\},  
  \end{equation}
  the spectra of $f^{(2)}_{ab}$ are represented by the set $\Lambda\cup\{4\}$.
 An example of the spectra is plotted in Figure~\ref{fig:spectra}.
 
\begin{figure}[h]
    \centering
    \includegraphics[width=.45\textwidth]{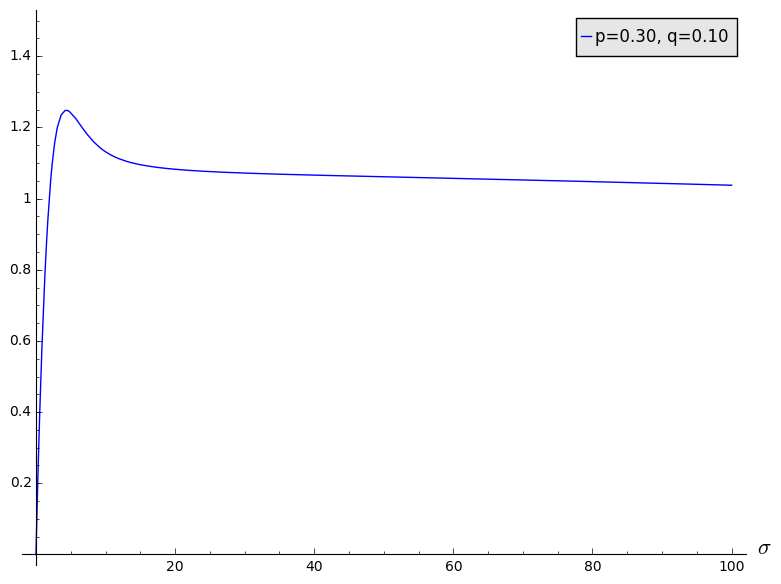}
    \includegraphics[width=.45\textwidth]{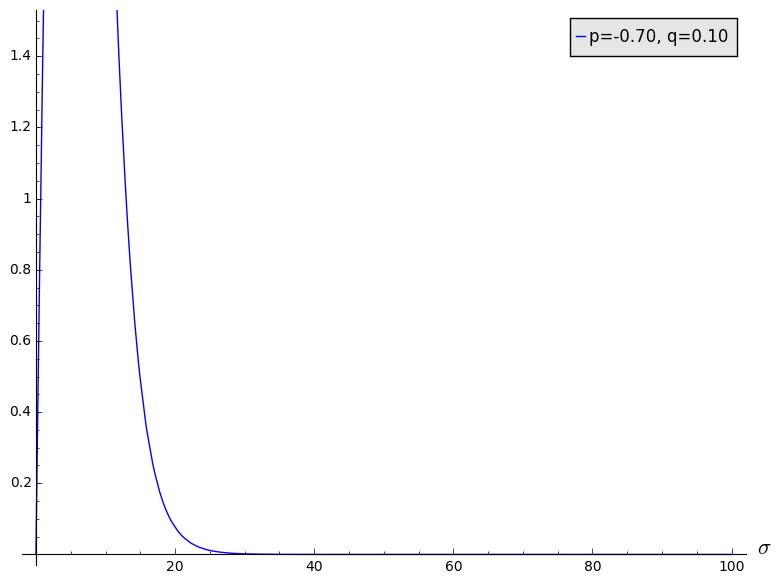}
    \caption{The behavior of $D(\sigma)$ is sensitive to the value of
      $\inf(\Lambda)$. The left one has its $2-4p-4|q|$ being
      positive and less than $4$, then everything is fine. The right
      one has its $2-4p-4|q|$ greater than $4$, and it is ill
      behaved.}   
    \label{fig:dim-exact}
  \end{figure}

As can be seen in the example, the spectra of $f^{(2)}$ are similar to those of $\nabla^2+m^2$ 
in one dimension, except that
the zero mode is supermassive and that higher spectra are deformed. 
Thus, the minimal of these spectra should be regarded as the rest mass, and an analogue of $\nabla^2$ 
on the emergent space can be evaluated by subtracting the mass from these spectra. 
Then, one way to understand the geometric property in general 
is to compute the spectral dimension defined by
  \begin{equation}
    \label{eq:dimensional}
    D(\sigma) \equiv -2 \frac{\sigma}{\Sigma} \frac{\partial
      \Sigma}{\partial \sigma} , 
  \end{equation}
  where
  \begin{equation}
    \label{eq:partial-function}
    \Sigma = e^{-(4-\lambda_{\text{min}})\sigma} +\sum_{\lambda \in \Lambda} e^{-(\lambda
      -\lambda_\text{min}) \sigma} ,
  \end{equation}
  and $\lambda_{\text{min}} = \min(\inf(\Lambda),4)$. For different $p$ and
  $q$, we can show numerical computations as examples, e.g.
  Fig~\ref{fig:dim-exact}. In the fine case of the figure, we certainly obtain 
  $D(\sigma)\sim 1$ in the large-scale region $\sigma  \gg 1$. 

  From the figure we realize that $4$ should not be the minimal,
  otherwise dimensions blow up as $N$ gets
  bigger. Noticing that the minimal of the set $\Lambda$ is given by
  $2-4p-4|q|$ when $N$ goes to infinity, one can immediately rewrite
  \begin{align}
    \lim_{N\rightarrow\infty} \frac{2\pi}{N} \Sigma 
    = & \lim_{N\rightarrow\infty} \sum_{n=1}^{N-1}\frac{2\pi}{N} \cdot
        \exp\left( -4|q| \sigma +4q \sigma \cos \frac{2\pi n}{N} \right)
        \notag \\  
    = & e^{-4|q|\sigma} \int_0^{2\pi} \text{d}\theta~ e^{4q \sigma \cos\theta} 
        = 2 \pi e^{-4|q|\sigma} I_0(4|q|\sigma) ,
    \end{align}
  where $I_0$ is the modified Bessel function of the first kind. Therefore
  the spectral dimension $D(\sigma)$ has an asymptotic behavior as
  \begin{align}
    \mathcal{D}_1(\sigma) \equiv \lim_{N\rightarrow\infty} D(\sigma) 
  = & - \frac{2\sigma}{e^{-4|q|\sigma}I_0(4|q|\sigma)} \frac{\text{d}}{\text{d} \sigma}
      \left(e^{-4|q|\sigma} I_0(4|q|\sigma)\right) \notag \\
  = & 2\tau \left(1 -\frac{I_1(\tau)}{I_0(\tau)}\right) ,
  \end{align}
  where $\tau=4|q|\sigma$.
  One can check that ${\cal D}_1(\sigma)$ behaves in the same manner as the example in the 
  left figure of Fig.~\ref{fig:dim-exact}.
  
The minimal of the spectrum should be no less than zero for the stability of the solution \eq{eq:flat1dim} under small perturbations.
 On the other hand, from the
  previous discussion, a proper result also requires the minimum be in the
  set $\Lambda$ instead of being $4$. Together they give the restriction of
  \begin{equation}
    \label{eq:restriction}
    0 \leqslant \inf(\Lambda) \leqslant 4 .
  \end{equation}
  When $N$ goes to infinity, this becomes
  \begin{align}
    & 0 \leqslant 2 -4p -4|q| \leqslant 4 \notag \\ 
    \label{eq:infty-restriction} \Rightarrow
    & 1 \geqslant 2p +2|q| \geqslant -1 ,
  \end{align}
  which obviously depends on the sign of $1+6\kappa+3\gamma$. Define
  $s=\text{sgn}(1+6\kappa+3\gamma)$. It is clear that
  $|1+6\kappa+3\gamma| = s(1+6\kappa+3\gamma) > 0$ thus $s^2\equiv1$. Then
  Eq.\eqref{eq:infty-restriction} can be rewritten as
  \begin{eqnarray}
    s\gamma & \geqslant & sf_s(\kappa) \equiv s(1 -2\kappa) +8 |\kappa| , \\
    s\gamma & \geqslant & sg_s(\kappa) \equiv -s\left( \frac{3}{5}
                          +2\kappa\right) -\frac{8}{5} |\kappa| , \\
    s\gamma & > & sh_s(\kappa) \equiv -s\left( \frac{1}{3} +2\kappa\right) .
  \end{eqnarray}
We understand the following:
  \begin{enumerate}
  \item{$s=+$:
  
    We have $f_+>h_+>g_+$ for any $\kappa$, thus the restriction is given by
    $\gamma \geqslant f_+$.}
  \item{$s=-$:   
  
    In area $|\kappa|>1/6,~f_-<h_-<g_-$; thus the restriction is $\gamma
    \leqslant f_-$. In area $|\kappa|<1/6,~f_->h_->g_-$; thus the restriction is
    $\gamma \leqslant g_-$. When $|\kappa|=1/6,~f_-=g_-=h_-$, however we know
    $\gamma$ should never equal to $h_-$. Therefore, the restriction is given by
    $\gamma<0$ and $\gamma<-2/3$ for $\kappa=-1/6$ and $\kappa=1/6$, respectively.}
  \end{enumerate}
  
\begin{figure}[h]
\begin{center}
\includegraphics[scale=1]{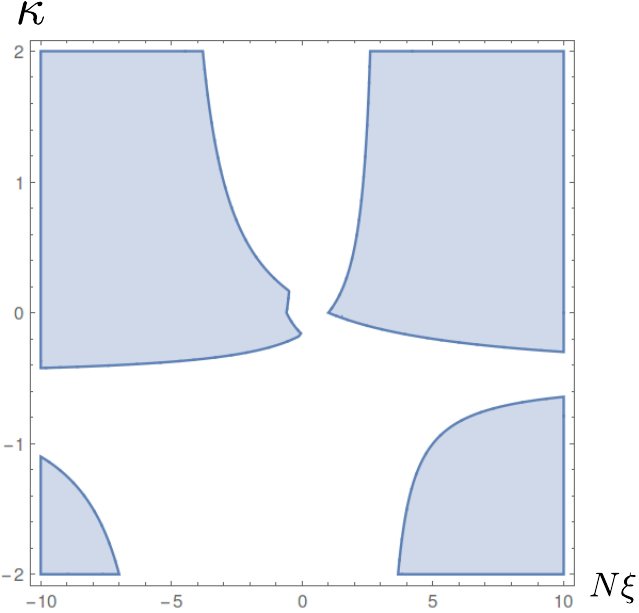}
\caption{The shaded regions represent the allowed region \eq{eq:infty-restriction}.}
\label{fig:restriction}
\end{center}
\end{figure}

  Taking the form of $\gamma=N\xi(1+2\kappa)$ into our final consideration,
 and noticing that the sign of $1+2\kappa$ also matters, 
 we give Table~\ref{table:restriction} and Figure~\ref{fig:restriction}
  describing the restriction in total.
  
  \begin{table}[h]
  \centering
    \begin{tabular}{c|c|c|c|c|c|c|c|c}
    \hline \multicolumn{2}{c|}{$\kappa$} 
    & $(-\infty,-\frac{1}{2})$ & $(-\frac{1}{2},-\frac{1}{6})$ 
    & $-\frac{1}{6}$ & $(-\frac{1}{6},0)$ & $(0,\frac{1}{6})$ 
    & $\frac{1}{6}$ & $(\frac{1}{6},+\infty)$ \\ \hline
    \hline \multirow{2}{*}{$N\xi\geqslant C_{u,v}(\kappa)$} & $u$
    & 3 & \multicolumn{3}{|c|}{-5} & \multicolumn{3}{|c}{3} \\
    \cline{2-9} & $v$ & -2
    & \multicolumn{3}{|c|}{6} & \multicolumn{3}{|c}{-2} \\
    \hline \multirow{2}{*}{$N\xi\leqslant C_{u,v}(\kappa)$} & $u$
    & -5 & 3  & $0^*$ & -9/5 & -1/5 & $-2/3^*$ & -5 \\
    \cline{2-9} & $v$
    & 6 & -2 & & 6/5 & -2/5 & & 6 \\ \hline
    \end{tabular}
  \caption{Here, $C_{u,v}(\kappa)=u+\frac{v}{1+2\kappa}$. The additional star
    mark means the equivalence to that restriction cannot be obtained.}
  \label{table:restriction}
  \end{table}

\subsection{Flat $D$-Dimensional Tori} 
\label{subsec:flatDdim}

  For simplicity, we still assume the localized $P^{L}$ is given by
  \begin{equation}
    \label{eq:P-localized-D}
    P^{L}_{I,I,I} = 1 \,,\quad
    P^{L}_{I,I,I+e^j} =
    P^{L}_{I,I+e^j,I+e^j} = \kappa_D ,
  \end{equation}
  where $I=(i_1,i_2,\dots,i_D)$ is a vector from a module defined on
  ring $\mathbb{Z}/\mathcal{N}\mathbb{Z}$, characterizing a point on the 
  $D$-dimensional compact lattice. $\mathcal{N}^D=N$ describes how fine the structure
  is divided. $e^j$'s are the basis of this module. In this case, if
  $J=I\pm e^j$, we say $J$ is a neighbor of $I$. Again, we denote
  this relation by $J \sim I$.  

  Generate the total tensor $P_{abc}$ via eq.\eqref{eq:configp}:
  \begin{eqnarray}
    \label{eq:P-specific-D}
    P_{I,I,I} & = & 1 +3\gamma_D , \\
    P_{I,J,J} & = &
                    \begin{cases}
                      \kappa_D +\gamma_D & J \sim I \\
                      \gamma_D & \text{otherwise}
                    \end{cases} , \\
    P_{I,J,K} & = & 0 ,
  \end{eqnarray}
  where $\gamma_D=\xi\sum_JP^L_{IJJ}=\xi(1+2D\kappa_D)$.

  Assuming that the field $\bar \phi$ in \eq{eq:flat1dim} with a uniform value on every point is the
  minimal of $f(P,\phi)$, we have
  \begin{align}
    P_{abc}\bar \phi_c = N^{-1/2} \sqrt{\frac{3}{2}} 
      &\left[(1 +3\gamma_D +2D(\kappa_D+\gamma_D) +(N-2D-1)\gamma_D)
        \delta_{ab} \right. \, \notag \\ 
      & +2(\kappa_D +\gamma_D) \tilde{\delta}_{ab} \left.+2\gamma_D (1_{ab} -\delta_{ab} -\tilde{\delta}_{ab})
        \right] \notag \\
    \label{eq:Pab-D}
    = N^{-1/2} \sqrt{\frac{3}{2}}
      & \left[(1 +2D\kappa_D +N\gamma_D)
        \delta_{ab} +2\kappa_D \tilde{\delta}_{ab} +2\gamma_D 1_{ab} \right],
  \end{align}
  and
  \begin{equation}
    \label{eq:Aphi-D}
    A(\bar \phi) = N^{-1/2}\left( \frac{3}{2} \right)^{3/2}\left(1 +6D\kappa_D +3N\gamma_D \right)
    . 
  \end{equation}
  Let us introduce
   $p,q,r$ in the same form as eq.\eqref{eq:pqr} with $\kappa=D\kappa_D$ and
  $\gamma=N\gamma_D$. We get 
  \begin{equation}
    \label{eq:free-energy-D}
    f^{(2)}_{ab}  = (2 -4p) \delta_{ab} -\frac{2q}{D}
    \tilde{\delta}_{ab} +\frac{6-4r}{N} 1_{ab} .
  \end{equation}
  
  Now consider vectors,
  $\Psi^{\hat{n}}_J = e^{i\frac{2\pi \hat{n}}{\mathcal{N}}\cdot{J}}$,
  where both $\hat{n}$
  and $J$ are $D$-component vectors. We can calculate that
  \begin{equation}
    \label{eq:eigen-vector-D}
    \sum_J f^{(2)}_{IJ} \cdot \Psi^{\hat{n}}_J = (2-4p)
    e^{i \frac{2\pi \hat{n}}{\mathcal{N}} \cdot I} -\frac{2q}{D} \sum_{j=1}^D
    \left(e^{i\frac{2\pi \hat{n}}{\mathcal{N}} \cdot (I +e^j)} +e^{i\frac{2\pi
          \hat{n}}{\mathcal{N}} \cdot (I -e^j)}\right) +\frac{6-4r}{N} \sum_J
    e^{i\frac{2\pi \hat{n}}{\mathcal{N}} \cdot J},
  \end{equation}
  where the second term on the right-hand side can be rewritten as
  \begin{equation}
    \label{eq:second-term}
    \frac{2q}{D} \sum_{j=1}^D
    \left(e^{i\frac{2\pi \hat{n}}{\mathcal{N}} \cdot (I +e^j)} +e^{i\frac{2\pi
          \hat{n}}{\mathcal{N}} \cdot (I -e^j)}\right) = \frac{4q}{D}
    e^{i\frac{2\pi \hat{n}}{\mathcal{N}} \cdot I} \sum_{j=1}^D
    \cos\frac{2\pi n_j}{\mathcal{N}} 
  . 
  \end{equation}
  The third term is
  \begin{align}
    \sum_J e^{i\frac{2\pi \hat{n}}{\mathcal{N}} \cdot J} =
      & \sum_{J} \prod_{k=1}^D e^{i\frac{2\pi n_k}{\mathcal{N}} J_k} =
        \prod_{k=1}^D \sum_{j=1}^\mathcal{N} e^{i\frac{2\pi n_k}{\mathcal{N}} j} \notag \\ 
    = &
        \begin{cases}
          N & \hat{n} = 0 \\
          0 & \text{otherwise}
        \end{cases} .
  \end{align}

  Thus, when $\hat{n}\neq0$, we get $N-1$ eigenvalues
  given by
  \begin{equation}
    \label{eq:eigen-value-D}
    f^{(2)}_{ab} \Psi^{\hat{n}}_b = \left( 2 -4p -\frac{4q}{D} \sum_{j=1}^D
      \cos\frac{2\pi n_j}{\mathcal{N}} \right) \Psi^{\hat{n}}_a .
  \end{equation}
  And when $\hat{n}=0$, we have
  \begin{equation}
    \label{eq:eigen-value-4}
    f^{(2)}_{ab} \Psi^{\hat{n}}_b = f^{(2)}_{ab}~1_b = 4 \cdot 1_a .
  \end{equation}
  Again, we have the spectra described as $(\Lambda_D \backslash \{2-4p-4q\})
  \cup \{4\}$, where
  \begin{equation}
    \label{eq:eigen-values-D}
    \Lambda_D = \left\{ \left. 2 -4p -\frac{4q}{D}\sum_{j=1}^D \cos
        \frac{2\pi n_j}{\mathcal{N}} \,\right|\, n_j = 0, 1, \dots, \mathcal{N}-1
  \right\} .
  \end{equation}

  Following the same process as in the one dimensional case, we have
  $0\leqslant  \lambda_{\text{min}} = 2-4p-4|q| \leqslant 4$. 
  The spectral dimension is still defined as \eqref{eq:dimensional}. 
  Then we
  have
  \begin{align}
    \lim_{\mathcal{N}\rightarrow\infty} \left( \frac{2\pi}{\mathcal{N}} \right)^D \Sigma
  = & \lim_{\mathcal{N}\rightarrow\infty} \left( \frac{2\pi}{\mathcal{N}} \right)^D \sum_{\lambda \in
      \Lambda_D} e^{(\lambda -\lambda_{\text{min}}) \sigma} 
      \notag \\  
  = & \lim_{\mathcal{N}\rightarrow\infty} e^{-4|q|\sigma} 
      \prod_{j=1}^D\, \sum_{\{n_j\}}\, \frac{2\pi}{\mathcal{N}} \exp\left( \frac{4q}{D} \sigma 
      \cos\frac{2\pi n_j}{\mathcal{N}} \right) \notag \\
  = & e^{-4|q|\sigma} \prod_{j=1}^D\, \lim_{\mathcal{N}\rightarrow\infty}
      \sum_{n=0}^{\mathcal{N}-1}\, \frac{2\pi}{\mathcal{N}} \exp\left(\frac{4q}{D} \sigma 
      \cos\frac{2\pi n}{\mathcal{N}} \right) \notag \\
  = & e^{-4|q|\sigma} \left[ 2 \pi I_0\left( \frac{4|q|}{D} \sigma \right)
      \right]^D .
  \end{align}
  Therefore, in $N\to\infty$, the spectral dimension behaves as
  \begin{equation}
    \label{eq:dim-asym-infty-D}
    \mathcal{D}(\sigma) \equiv \lim_{N \rightarrow \infty} D(\sigma) = 
    D \cdot \mathcal{D}_1(\sigma/D) .
  \end{equation}
  
  Since, $\lambda_\text{min}$ and $p,q,r$, all takes the same form as in
  the one dimensional case, the restriction condition for $\xi$
  listed in Table~\ref{table:restriction} still applies with $\kappa=D\kappa_D$. 

\section{Spaces with general metrics}
\label{sec:general}

In Section \ref{sec:flat}, we have shown that 
the two-point correlation function, 
\textit{i.e.} $f^{(2)-1}_{ab}$, 
indicates the emergence of Euclidean flat spaces on boundaries of networks. 
Putting the argument forward, we try to read off metrics for more general emergent
spaces, when $N$ is large. 
To begin, we parametrize $P_{abc}$ in the same way as in \eqref{eq:P-local} 
with more general $P^L_{abc}$, where 
$P^L_{abc}$ is an almost local part of the tensor such that $P^L_{abc} \ne 0$ only for $a\sim b\sim c$ (In this section, the symbol $\sim$ is used to represent
nearby points as well as the same point, 
slightly more general than the usage in Section~\ref{sec:flat}.).
The strategy is that we will compute the quadratic term of $v$ in \eqref{eq:partbyy}
to extract a metric in an effective field theory in a formal continuum limit.

To set aside a super-massive mode similar to the zero mode for the flat case 
in Section~\ref{sec:flat},  
we assume perturbations $v$ around the vacuum $\bar \phi$ to satisfy 
a constraint, 
\[
v_a \bar \phi_a =0.
\]
This is a non-local kind of constraint, since, 
in the present general case too, we expect that 
$\bar \phi_a$ takes non-vanishing values for all the range of $a$.
This non-local constraint would not ruin the significance of our model, 
since it would be impossible for a local observer to detect 
the existence of such a single non-local constraint, when $N$ is large.
Then the contraction of $v_a$ with the Hessian \eqref{eq:f2ab} is given by
\[
f^{(2)}_{ab}v_av_b 
=
\left( 
2- \frac{6 \xi  }{A(\bar \phi)} P^L_{bdd} \bar \phi_b
\right) v_a v_a 
-\frac{6}{A (\bar \phi)} P^L_{abc} \bar \phi_a v_b v_c. 
\label{eq:f2vv}
\]

When $N \to \infty$, we implement the formal continuum limit:
\[
P^L_{abc} \Rightarrow  P^L(x,y,z), 
\ \ \ 
\bar{\phi}_a \Rightarrow  \bar{\phi} (x), 
\ \ \ 
v_a \Rightarrow  v(x), 
\ \ \ 
\sum_a \Rightarrow \int \text{d}^Dx,
\label{eq:continuum}
\]
where the discrete labels $a,b,c$ have been replaced by the continuous coordinates 
$x,y,z \in \mathbb{R}^D$, respectively. 
Although $P^L(x,y,z)$ is a tri-local function, by definition, $P^L(x,y,z) = 0$ unless $x \sim y \sim z$. 
In other words, $P^L(x,y,z)$ can be regarded as a distribution concentrated around
$x \sim y \sim z$, and a moment expansion, which was introduced in another context 
in \cite{Sasakura:2015pxa}, will give a good approximation:
\[
&\int \text{d}^Dy\text{d}^Dz\ 
P^L(x,y,z) f_1(x) f_2(y) f_3(z) \notag \\
& \ \ \  \cong
\alpha f_1f_2f_3 
+\beta^{\mu} f_1 \partial_{\mu} 
\left( f_2f_3 \right)
+\frac{1}{2} \gamma^{\mu \nu} f_1
\left( f_2 \partial_{\mu}\partial_{\nu} f_3 + f_3 \partial_{\mu}\partial_{\nu} f_2  \right) 
+ \tilde \gamma^{\mu \nu} f_1 \partial_{\mu}f_2 \partial_{\nu} f_3 
+ \cdots,
\label{eq:momentexp}
\]
where
$f_1$, $f_2$ and $f_3$ are test functions, 
$\alpha$, $\beta^{\mu}$, $\gamma^{\mu \nu}$ 
and $\tilde \gamma^{\mu \nu}$ are the moments,  
the argument $x$ is suppressed in the last line for brevity,  
and the dots represent higher moments being neglected in the present analysis.  
Due to the symmetry of $P^L(x,y,z)$ under the exchange of its arguments $x,y,z$, 
the moments are not independent and satisfy \cite{Sasakura:2015pxa} 
\[
\beta^{\mu} = \frac{1}{2} \partial_{\nu} \gamma^{\mu \nu},
\ \ \ 
\tilde \gamma^{\mu \nu} = \frac{1}{2} \gamma^{\mu \nu}.
\label{eq:betagamma}
\]

As has been shown in \cite{Sasakura:2015pxa},
the O($N$) invariance \eq{eq:orthogonal} of the randomly connected tensor network
implies that the formal continuum limit of the system be invariant under 
the spatial diffeomorphism.
The spatial diffeomorphism is such that, after the formal replacement 
$a\Rightarrow x$, a vector, say $f_a$, becomes a scalar half-density $f(x)$.
Therefore the test functions above should be treated as 
scalar half-densities, and 
the moments defined above are not covariant.  
In order to make the transformation property transparent, 
let us introduce a symmetric two-tensor $g_{\mu \nu}$, which will shortly be related to 
the moments,
and rewrite the expansion (\ref{eq:momentexp}) as
\[
&\int \text{d}^Dy\text{d}^Dz\ 
P^L(x,y,z) f_1(x) f_2(y) f_3(z) \notag \\
& \ \ \  \cong
\alpha_c f_1f_2f_3 
+\beta^{\mu}_c f_1 \nabla_{\mu} 
\left( f_2f_3 \right)
+\frac{1}{2} \gamma^{\mu \nu}_c f_1
\left( f_2 \nabla_{\mu} \nabla_{\nu} f_3 + f_3 \nabla_{\mu}\nabla_{\nu} f_2  \right) 
+ \tilde \gamma^{\mu \nu}_c f_1 \nabla_{\mu}f_2 \nabla_{\nu} f_3 
+ \cdots,
\label{eq:momentexpcovariant}
\]
where  $\nabla_{\mu}$ is the covariant derivative 
associated with $g_{\mu \nu}$, 
and $\alpha_c$, $\beta^{\mu}_{c}$ and $\gamma^{\mu \nu}_c$ are 
the covariant moments.
Below, we will shortly see that $g_{\mu \nu}$ actually plays the role of a \textit{metric} in an effective action. 
Corresponding to (\ref{eq:betagamma}),  we also have 
\[
\beta^{\mu}_c = \frac{1}{2} \nabla_{\nu} \gamma^{\mu \nu}_c, 
\ \ \ 
\tilde \gamma^{\mu \nu}_c = \frac{1}{2} \gamma^{\mu \nu}_c.    
\label{eq:betagammacovariant}
\] 
As shown in Appendix \ref{sec:covariance}, 
the covariant moments are just given by the linear combinations of the moments above
up to the higher order corrections neglected in the present analysis,
and therefore, the two descriptions, \eq{eq:momentexp} and \eq{eq:momentexpcovariant},
are equivalent up to the order.

By using these covariant properties, 
it is possible to rewrite the continuum limit of $v$, $\bar \phi$ and 
the covariant moments, $\alpha_c,\gamma^{\mu \nu}_c$, in the following covariant manner: 
\[
v = g^{1/4} \sigma_1 , 
\ \ \ 
\bar \phi = g^{1/4} \sigma_2, 
\ \ \ 
\alpha_c = g^{-1/4} \sigma_3, 
\ \ \ 
\gamma^{\mu \nu}_c = g^{-1/4} g^{\mu \nu} \sigma_2, 
\label{eq:weights}
\]
where we have introduced scalars, $\sigma_i\ (i=1,2,3)$,
and $g$ is the determinant of the metric $g_{\mu \nu}$. 
Here, the vectors, $v$ and $\bar \phi$,
are rewritten as scalar half-densities, and the weights of $\alpha_c$ and  $\gamma_c$
are determined from the invariance of \eq{eq:momentexpcovariant}.
Note also that we have related $g^{\mu\nu}$, which was initially arbitrary, with $\gamma^{\mu\nu}$
in a covariant manner, and the arbitrariness of 
the Weyl rescaling of $g_{\mu\nu}$ has been fixed so that 
\footnote{In \cite{Sasakura:2015pxa},
the Weyl rescaling of the metric is chosen differently with another reasoning in such a way as to satisfy 
$\alpha_c \gamma^{\mu \nu}_c = g^{-1/2} g^{\mu \nu}$.}   
\[
\frac{1}{\bar \phi} \gamma^{\mu \nu}_c 
= g^{-1/2}g^{\mu \nu}.
\label{eq:phigamma}
\]

Let us apply the formal continuum limit (\ref{eq:continuum}) to \eq{eq:f2vv},
using (\ref{eq:momentexpcovariant}) and \eq{eq:weights}. The second term 
becomes  
\[
P^L_{abc}\bar \phi_a v_b v_c 
&
\Rightarrow
\int \text{d}^Dx \text{d}^Dy \text{d}^Dz\, 
P^L (x,y,z) \bar \phi(x)v(y)v(z) \notag \\
&=\int \text{d}^Dx
\left( 
\alpha_c \bar \phi v^2 
+\bar \phi (\nabla_\nu \gamma^{\mu\nu}_c) v \nabla_{\mu} v 
+\frac{1}{2}\bar \phi  \gamma^{\mu \nu}_c (\nabla_{\mu}v)( \nabla_{\nu}v)
+\bar \phi v \gamma_c^{\mu\nu} \nabla_\mu \nabla_\nu v   
+ \cdots
\right)  \notag \\
&= \int \text{d}^Dx 
\left(
\alpha_c \bar \phi v^2 
- \frac{1}{2} \bar \phi \gamma^{\mu \nu}_c \nabla_{\mu} v \nabla_{\nu} v 
- v \gamma^{\mu \nu}_c \nabla_{\mu}v \nabla_{\nu} \bar \phi 
+ \cdots
\right)  \notag \\
&= \int \text{d}^Dx\ \sqrt{g} 
\left(
\sigma_1{}^2 \sigma_2\sigma_3 
- \frac{1}{2} \sigma_1{}^2 g^{\mu \nu}\nabla_{\mu}\sigma_1 \nabla_{\nu}\sigma_2 
- \sigma_1\sigma_2 g^{\mu \nu} \nabla_{\mu} \sigma_1 \nabla_{\nu}\sigma_2
+ \cdots
\right)  \notag \\
&= \int \text{d}^Dx\ \sqrt{g} 
\left[
-\frac{1}{2}g^{\mu \nu} \nabla_{\mu}\Phi \nabla_{\nu} \Phi
+ 
\left(
e^{\lambda} 
+ \frac{1}{8} g^{\mu \nu} \nabla_{\mu} \rho \nabla_{\nu} \rho
\right)
\Phi^2
+ \cdots
\right]. 
 \label{eq:pphivv}
\]
In the last line we have rewritten the scalars as
\[
\Phi := g^{-1/2}\bar \phi v = \sigma_1 \sigma_2, \ \ \ 
e^{-\rho/2} := \sigma_2, \ \ \ 
e^{\lambda} := \frac{\sigma_3}{\sigma_2}.  
\label{eq:scalars}
\] 
Taking into account the argument above, 
the whole part becomes
\[
f^{(2)}_{ab}v_av_b 
\Rightarrow \frac{6}{A}S + \cdots,
\label{eq:wholecontinuum}
\] 
where $S$ is given by
\[
S=
\int \text{d}^Dx\, 
\sqrt{g}
\left[
\frac{1}{2} g^{\mu \nu}\nabla_{\mu} \Phi \nabla_{\nu} \Phi 
-
\left\{
e^{\lambda} 
+ \frac{1}{8} g^{\mu \nu}\nabla_{\mu} \rho \, \nabla_{\nu} \rho 
-
\left( 
\frac{A}{3} - \xi B^L
\right) e^{\rho}
\right\} \Phi^2
\right],
\label{eq:action}
\]
with $A$ and $B^L$ being constants defined by
\[
&A :=
\int \text{d}^Dx \text{d}^Dy\text{d}^Dz\, 
P(x,y,z) \bar \phi (x) \bar \phi (y) \bar \phi (z), 
\label{eq:acont} \\
&B^L := 
\int \text{d}^Dx \text{d}^Dy\, 
P^L(x,y,y) \bar \phi (x).
\label{eq:blcont}
\]

$S$ is considered to be an effective action for the perturbation 
$\Phi$, while $\rho$, $\lambda$ and $g^{\mu\nu}$ are the background fields
determined by $\alpha_c$, $\gamma_c$ and $\bar \phi$.
The action is valid when the spatial dependence of the variations of 
the fields are small.   
The action has certainly a covariant form, which is an outcome of the underlying O($N$)
invariance of the system and the consistency of the reparametrization \eq{eq:weights}.
Lastly, we stress that the metric originates from the ``soft" non-local effects of $P^L$, 
characterizing the derivative terms in the action.

\section{Breakdown of the flat space} 
\label{sec:phasetransition}
In Section~\ref{sec:flat}, the $P$ which realizes the flat space
contains a non-local part proportional to $\xi$. 
Generally speaking, such a non-local part would cause troubles by breaking
the locality of a system, and would make the system unrealistic. 
In the present case, however, the spectra of the perturbations around the background 
agree with those of a usual scalar field theory in a flat space, 
except for the zero mode.
This single anomalous behaviour of  the zero mode will be 
ignorable for a local observer in the large $N$ limit, and 
therefore our model can be considered realistic.  
In fact, the non-local part of $P$ is indispensable for 
the stability of the flat space: 
as can be checked easily in \eq{eq:eigen-values-D}, if $\xi=0$, the Hessian matrix contains negative eigenvalues,
and the flat space is not a stable local minimum. 
Thus, if we start from a locally stable flat space with a sufficiently large $|\xi|$, 
and reduce it continuously, 
the flat space will eventually become unstable and decay into a new configuration.
This is a breakdown of the flat space. If this picture of the transition being triggered by 
appearance of negative eigenvalues is right, 
this process will be a second order phase transition.
However, this is not always true. 
In fact, in a certain parameter region of $\kappa,\xi$, 
there exists a distinct stable configuration which has a free energy lower than
that of the flat space.
In this region, the flat space is not a global minimum, and decays to it 
generally with a first-order phase transition.   
As we will see shortly, this new configuration describes a {\it point-like} space.   

For simplicity, we will restrict the following discussions to the one-dimensional case, 
the circles, 
though similar phenomena can also be expected for the $D$-dimensional tori and 
the curved cases.  
Let us first show, by a qualitative estimation, that such a point-like profile of ${\bar \phi}$ 
can actually have a free energy lower than that of the flat space.  
Let us consider the two configurations described by
\[
\begin{split}
{\bar \phi}_a^{flat}&=\sqrt{\frac{3}{2N}},\\
{\bar \phi}_a^{point}&=\sqrt{\frac32}\delta_{a1}.
\end{split}
\label{eq:compareconf}
\]
The former describes the flat space in Section~\ref{sec:flat}, 
and the latter describes a point-like space,
as it has a non-vanishing component only at, say, $a=1$. The normalization is taken 
so that they satisfy \eqref{eq:phi2}. Note that 
the latter configuration does not have to satisfy the stationary equation,
because we are just interested in whether there exists a profile of ${\bar \phi}$ 
with a lower free 
energy than that of the flat space. Because of the normalization ${\bar \phi}^2=\frac{3}{2}$, the free energy can be 
compared by the values of $|A({\bar \phi})|$, as can be seen in \eq{eq:freeenergywithA}.
By substituting \eq{eq:compareconf} to $A({\bar \phi})$, one obtains 
\[
\begin{split}
A({\bar \phi}^{flat})&=\left( \frac{3}{2} \right)^\frac{3}{2}\left( \frac{1+6\kappa}{\sqrt{N}}+3 \sqrt{N} (1+2 \kappa)\xi \right), 
\\
A({\bar \phi}^{point})&=\left(\frac{3}{2}\right)^\frac{3}{2}\left( 1+3 (1+2 \kappa)\xi \right).
\end{split} 
\]
When $N \gg 1$, which is the case of our main  interest,  $|A({\bar \phi}^{flat})|<|A({\bar \phi}^{point})|$ for $|\xi| \lesssim \frac{1}{\sqrt{N}}$, and hence ${\bar \phi}^{point}$ has a lower free energy.
Therefore, the flat space is globally stable only at $|\xi|\gtrsim \frac{1}{\sqrt{N}}$.  
On the other hand, the second order phase transitions
triggered by negative eigenvalues should occur at $|\xi| \sim \frac{1}{N}$,
because the eigenvalues are functions of $\gamma \sim N \xi$ as can be seen in \eq{eq:spectra}.
Since the estimated location $|\xi|\sim \frac{1}{N}$ is outside the globally stable region of 
the flat space, the transition between the flat and a point-like space 
is more likely to occur than the second order phase transition.
Since the two configurations in \eq{eq:compareconf} are largely 
different from each other, 
the phase transition between the flat and the point-like spaces should be first 
order in general.

\begin{figure}
\begin{center}
\includegraphics[scale=.4]{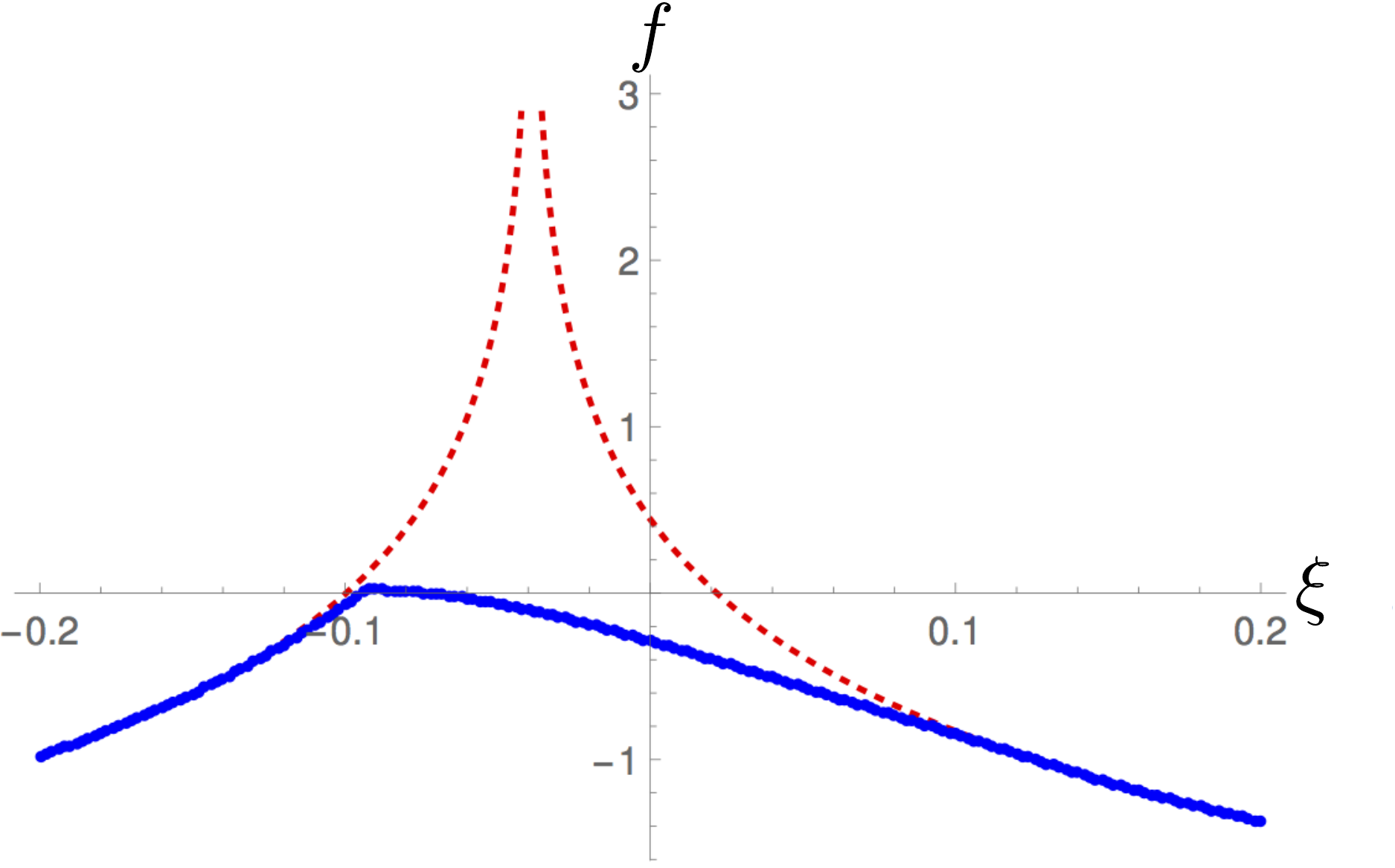}
\includegraphics[scale=.4]{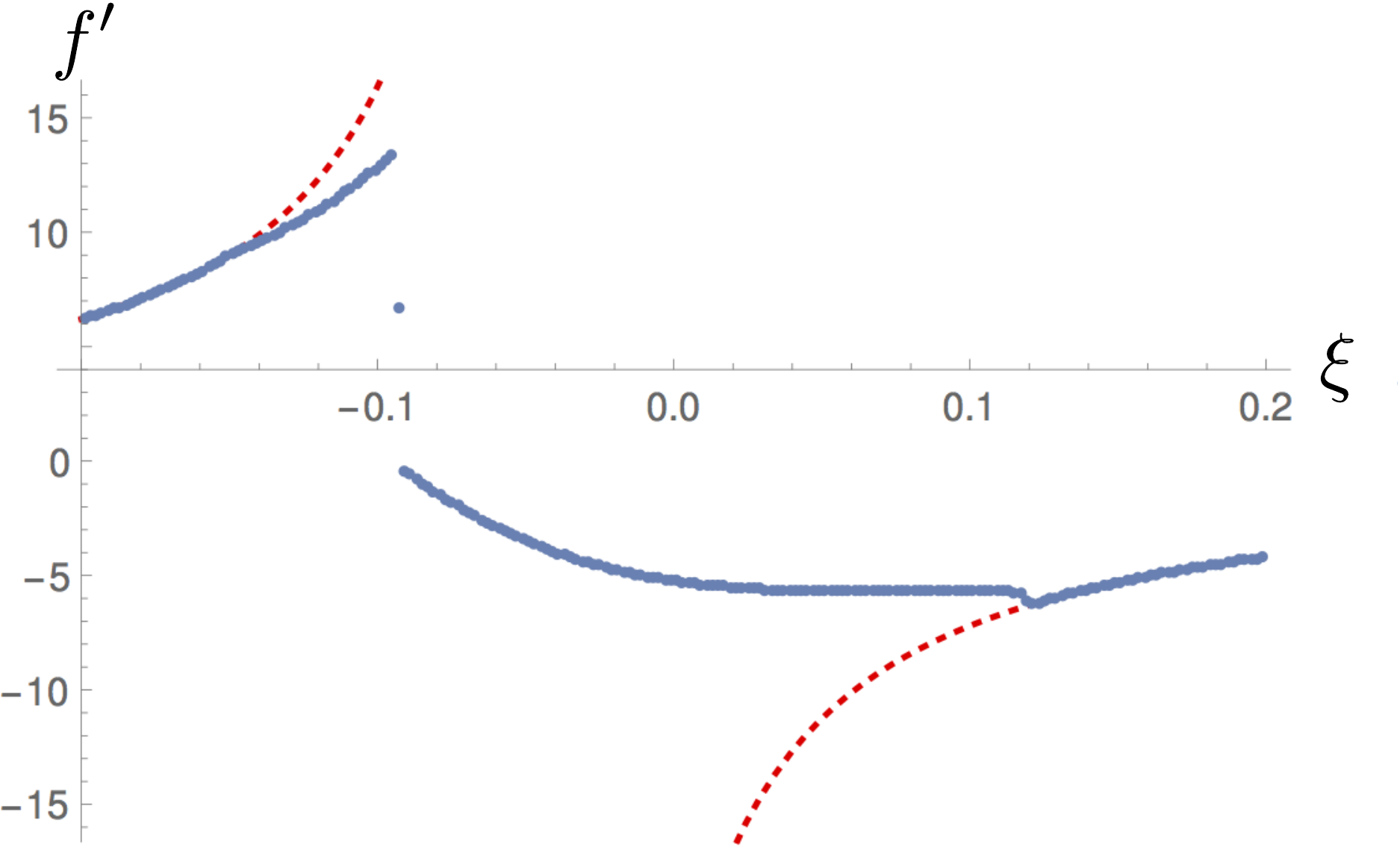}
\caption{The left: the free energy for $N=20,\ \kappa=1$. The horizontal and vertical axes
represent $\xi$ and the free energy, respectively.
The right: the first derivative of the free energy obtained by taking the finite 
differentials of the data in the left figure. 
In the two figures, the dotted lines represent the free energy and its first 
derivative of the flat space.
\label{fig:withk=1}
}
\end{center}
\end{figure}

The following numerical results support the qualitative discussion above. In fact, 
the actual phase structure is more complex than that. 
We have considered some fixed values of $N,\ \kappa$ and have gone through  discrete 
values of $\xi$ with small intervals.  
For every $\xi$, a numerical search for the global minimum of 
the free energy has been performed, and the minimum value is plotted.
The result for $N=20,\ \kappa=1$ is shown in the left figure of Fig.~\ref{fig:withk=1}.
We also evaluate the first derivative of the free energy with respect to $\xi$
by taking the finite differentials of these data, and the result is plotted in the
right figure of Fig.~\ref{fig:withk=1}. 
On the other hand,
the free energy of the flat space can be computed analytically by using the results 
in Section~\ref{sec:flat}, and is plotted by the dotted lines for the comparison. 
If a data point is apart from the dotted lines, it is in a phase different from the flat space. 
From these figures, we can see that 
there seem to exist four phases, and we label them by their regions:   
(i) $\xi \gtrsim 0.12$,
(ii) $-0.1\lesssim \xi \lesssim 0.12$, 
(iii)$-0.14 \lesssim \xi \lesssim -0.1$,
(iv) $\xi\lesssim -0.14$.
From the right figure, one can see that the phase transitions between (i) and (ii) and between (ii) and (iii) are first-order (the former is much weaker), while that between (iii) and (iv) is second-order.  

\begin{figure}
\begin{center}
\includegraphics[scale=.4]{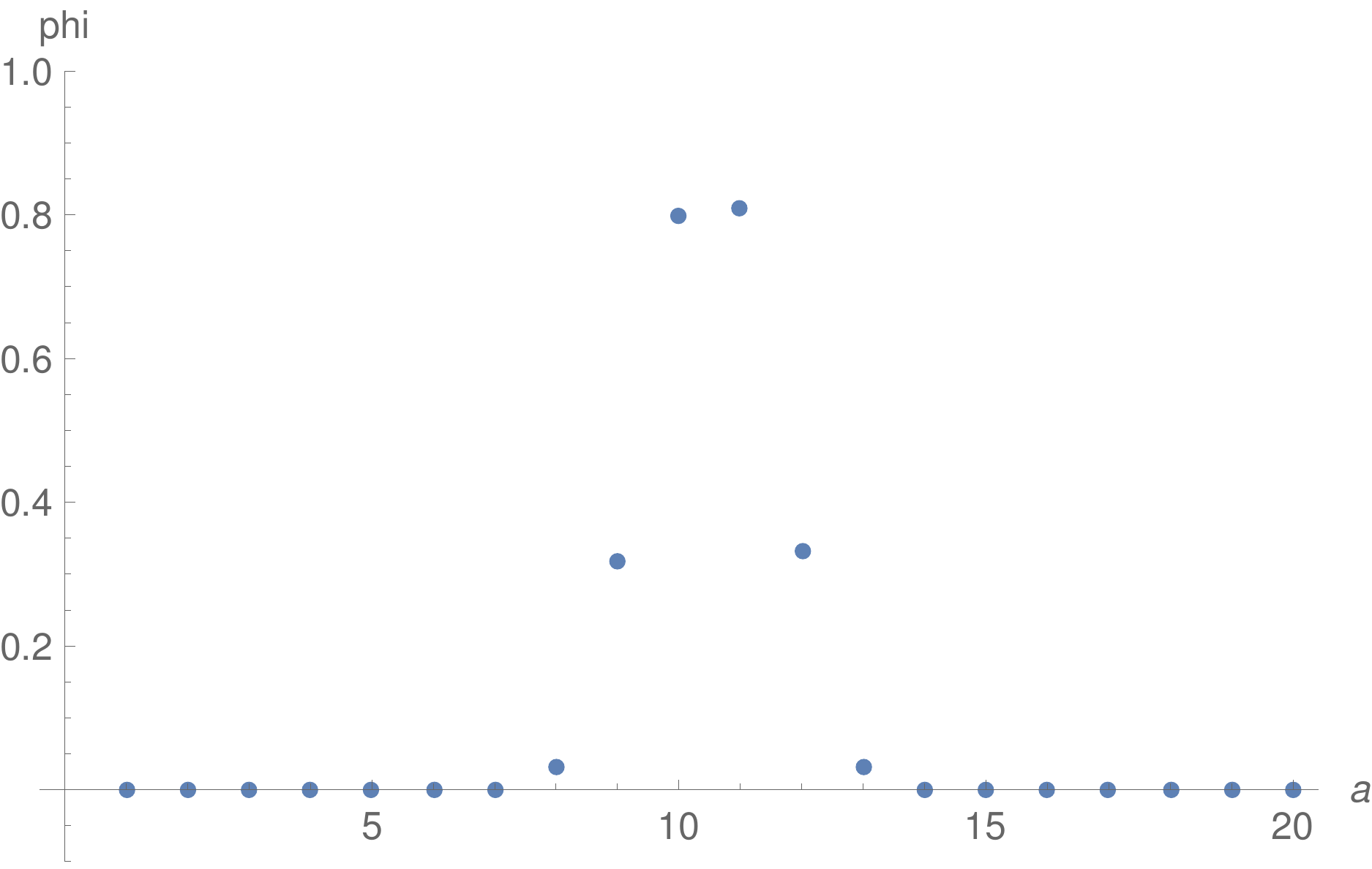}
\includegraphics[scale=.4]{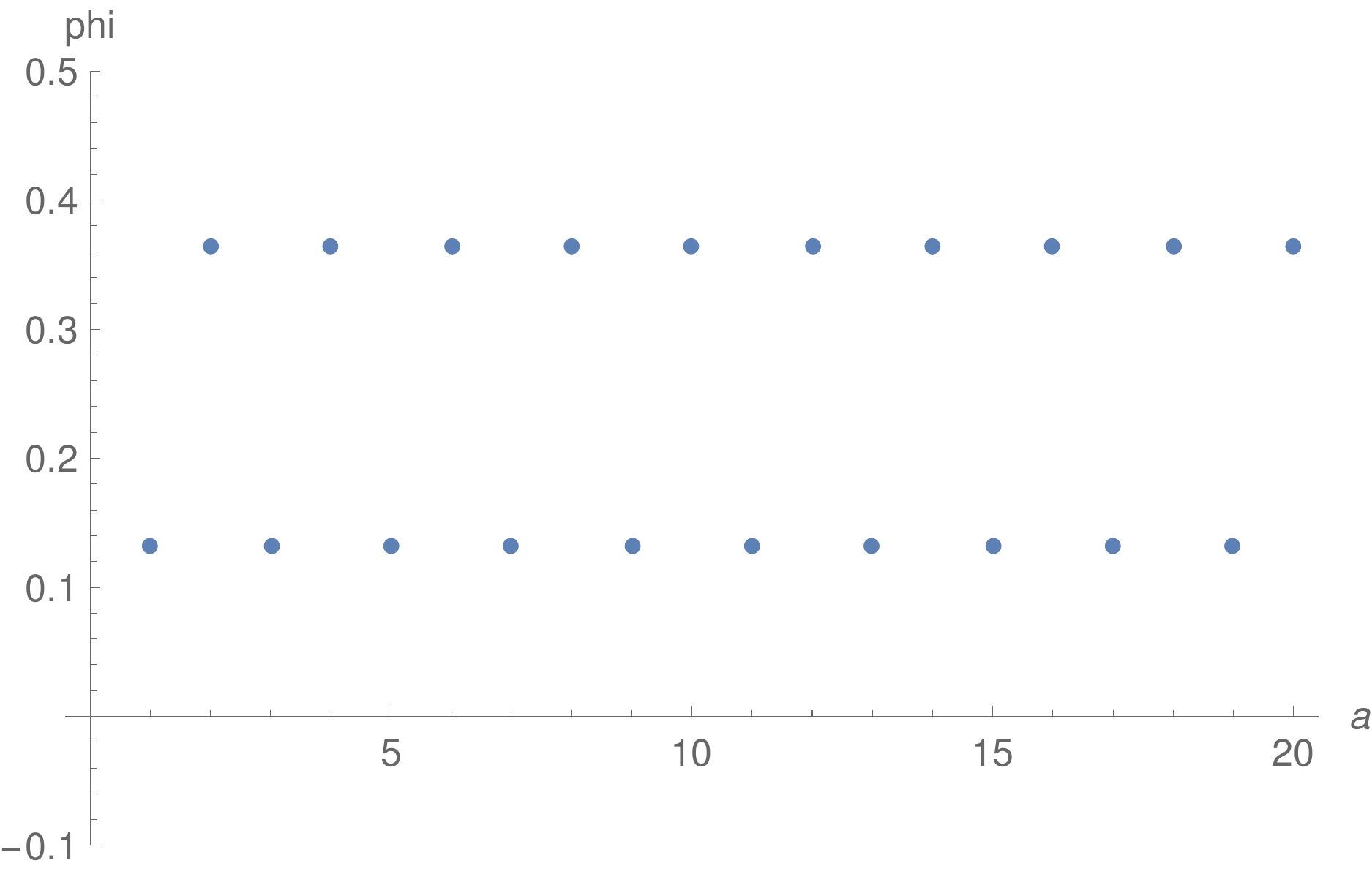}
\caption{The configurations of ${\bar \phi}$ in the phase (ii) ($\xi=0$) 
and (iii) ($\xi=-0.12$) for $N=20,\ \kappa=1$ 
are plotted in the left and right figures, respectively.
The vertical and horizontal axes represent the values of ${\bar \phi}_a$ and the 
index $a$, respectively. 
}
\label{fig:confphik=1}
\end{center} 
\end{figure}

In the phase (i) and (iv), ${\bar \phi}$ takes the constant value $\sqrt{\frac{3}{2N}}$, 
and they represent the flat space in Section~\ref{sec:flat}.
They are certainly in the regions with large $|\xi|$,
which is consistent with the qualitative discussion above.
As can be seen in the left figure of Fig.~\ref{fig:confphik=1}, 
in the phase (ii), ${\bar \phi}$ takes non-zero values only
around a certain point ($a \sim 10$ in the figure). This is the phase of a point-like space discussed above.
Because $P$ is invariant under the discrete translation 
$a\rightarrow a+l\,{\rm mod}\,  N$ with arbitrary integers $l$, 
the location of the central point is arbitrary. 
This is a phase where the discrete translation symmetry is all broken, since
${\bar \phi}_a\neq {\bar \phi}_{a+l\,{\rm mod }\, N}$. 
On the other hand, the right figure of Fig.~\ref{fig:confphik=1} shows that,
in the phase (iii), ${\bar \phi}$ oscillates with period 2. 
Therefore, ${\bar \phi}$ remains invariant under the even translations, ${\bar \phi}_a= {\bar \phi}_{a+2l\,{\rm mod }\, N}$,
and this is a phase where the discrete translation symmetry is only partially broken.  
The second order of the phase transition between (iii) and (iv) suggests that 
the transition is triggered by a negative eigenvalue. 
In fact, from the result of Section~\ref{sec:flat}, one can easily get 
that this is triggered by the change of the sign of 
the eigenvalue of the mode with $n=\frac{N}{2}$.
The mode is an oscillatory mode with period 2,
and the oscillatory profile of ${\bar \phi}$ in the phase (iii) is  explained by  the condensation of the mode.  

\begin{figure}
\begin{center}
\includegraphics[scale=.7]{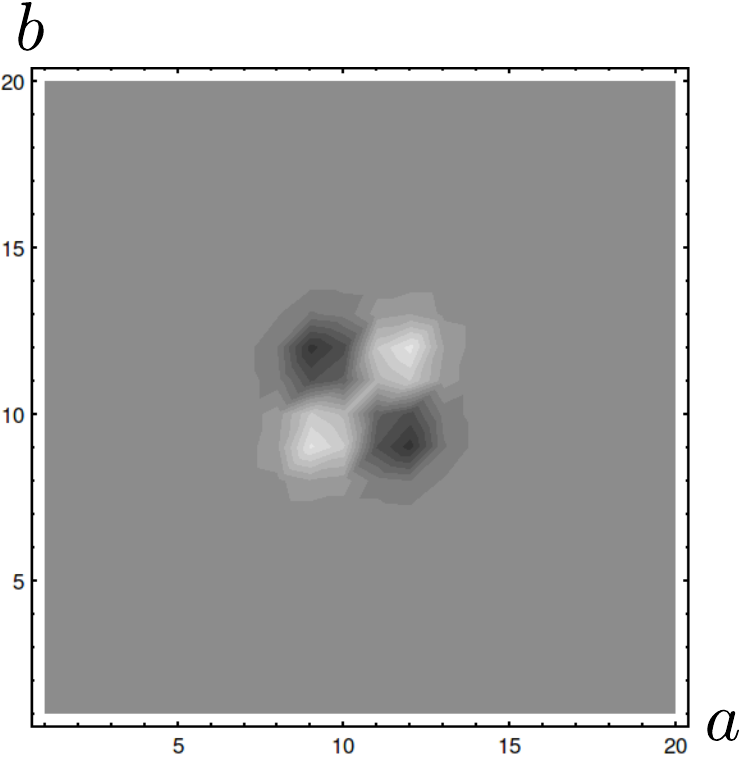}
\caption{The contour plot of the connected two-point function 
$\langle\varphi_a \varphi_b \rangle_\text{con.}$
for $N=20,\ \kappa=1,\ \xi=0$,
corresponding to the case with ${\bar \phi}$ in 
the left figure of Fig.~\ref{fig:confphik=1}.
The white and black regions represent positive and negative values, respectively,
while the gray region represents vanishing values.
Thus, in the point-like space, 
the non-vanishing values of the correlation function are concentrated only 
around $a,b\sim 10$, namely the non-vanishing region of $\bar \phi$.
}
\label{fig:k=1cor}
\end{center} 
\end{figure}

Let us explain the reason why the phase (ii) represents a point-like space. 
One reason is that ${\bar \phi}$ takes non-vanishing values 
only in a small point-like region, 
as around $a\sim 10$ in the left figure of Fig.~\ref{fig:confphik=1}.  
A physically more convincing explanation is that the correlation function takes 
non-vanishing values only in the vicinity of the non-vanishing region of ${\bar \phi}$. 
Fig.~\ref{fig:k=1cor} plots the two-point correlation function
for the phase (ii) ($\xi=0,\ N=20,\ \kappa=1$) for the case of ${\bar \phi}$ in the left figure of Fig.~\ref{fig:confphik=1}. 

\begin{figure}
\begin{center}
\includegraphics[scale=.4]{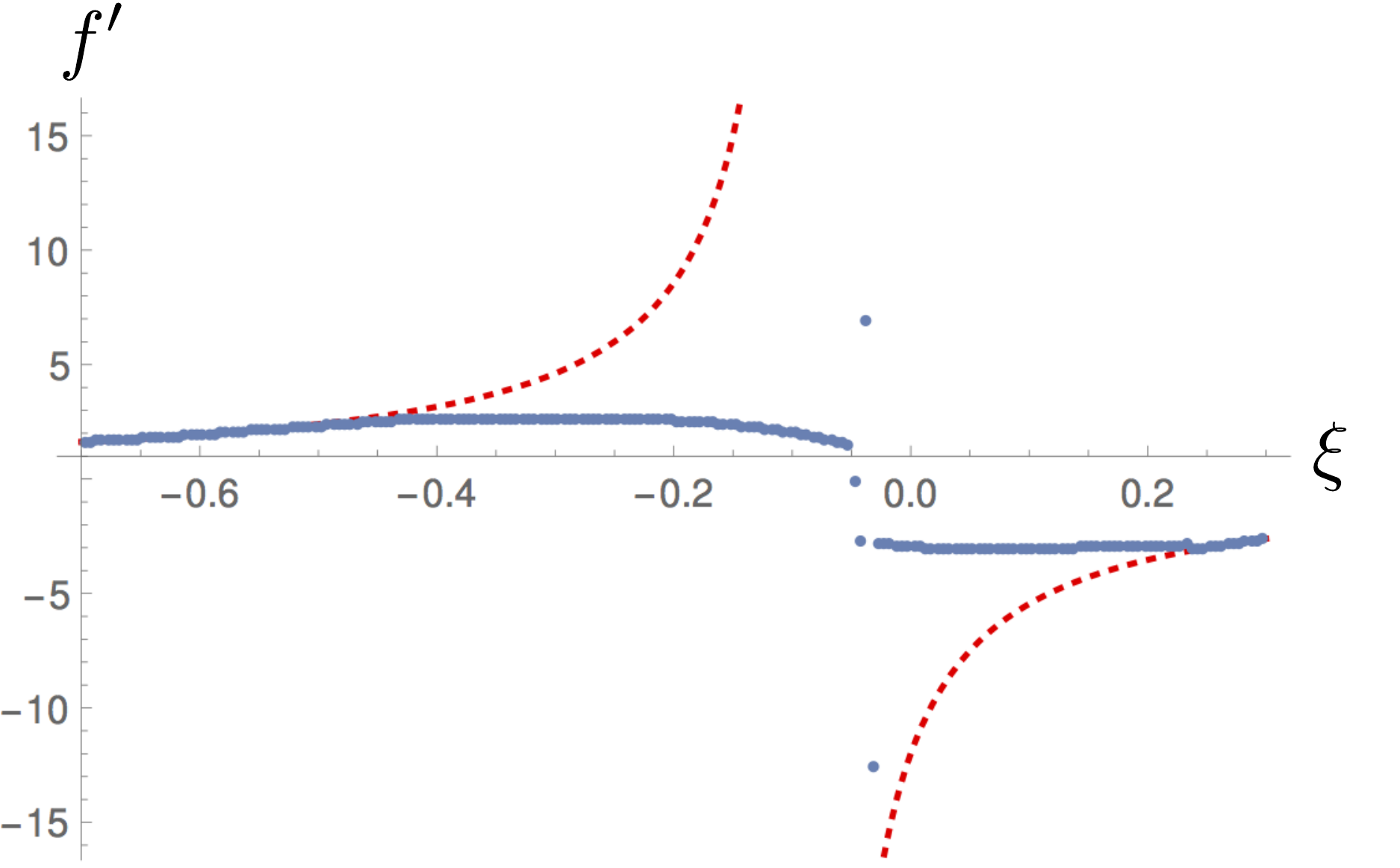}
\includegraphics[scale=.4]{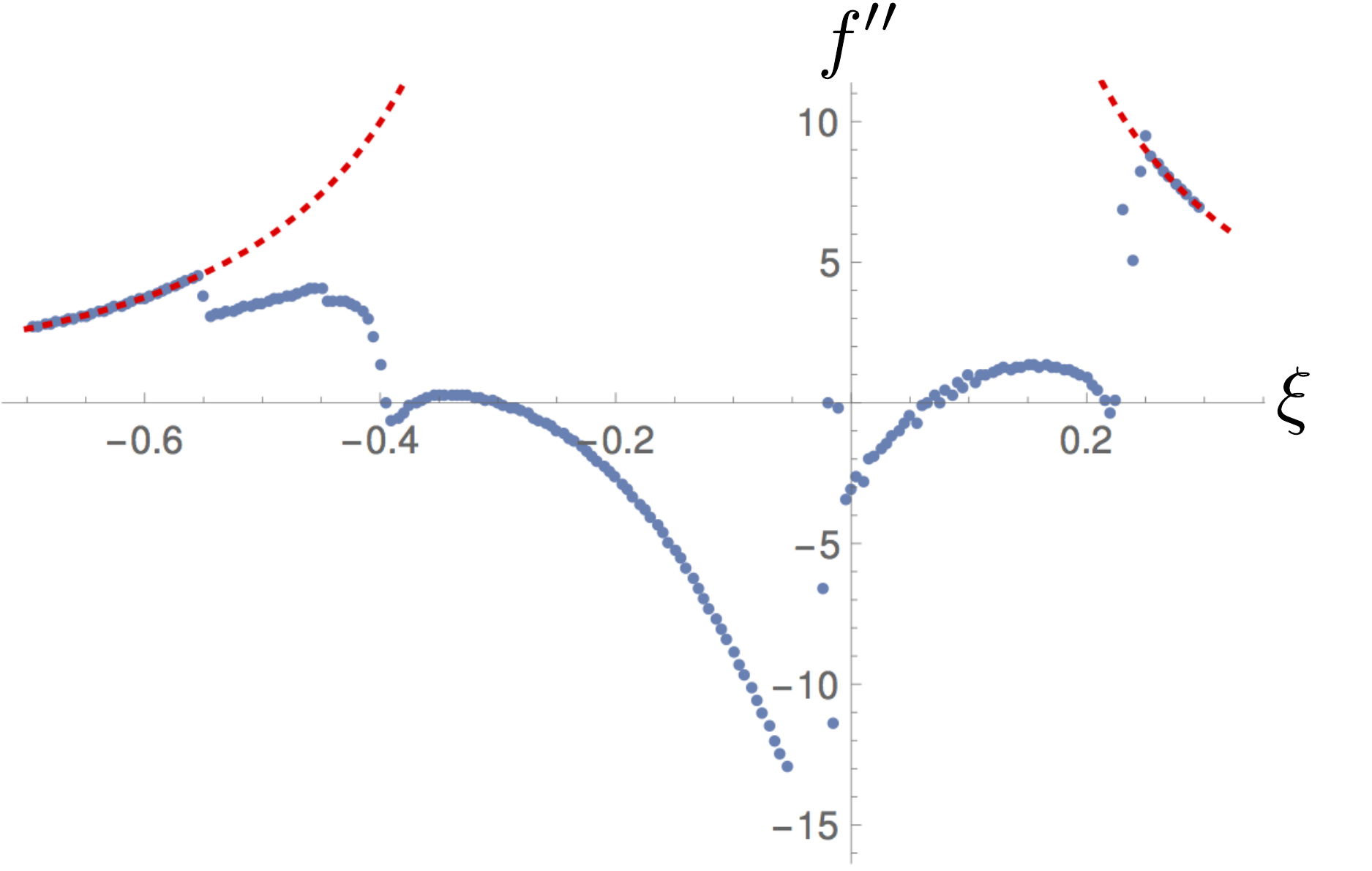}
\caption{The first (left) and second (right) derivatives of the free energy
for $N=20,\ \kappa=-1$. The horizontal axis represents $\xi$.
The dotted lines represent the case of the flat space. 
There seem to exist two first order phase transitions at the similar locations  
as in the previous case with $\kappa=1$, while there seem to exist more second order transitions.
}
\label{fig:freek=-1}
\end{center} 
\end{figure}

\begin{figure}[!h]
\begin{center}
\includegraphics[scale=.35]{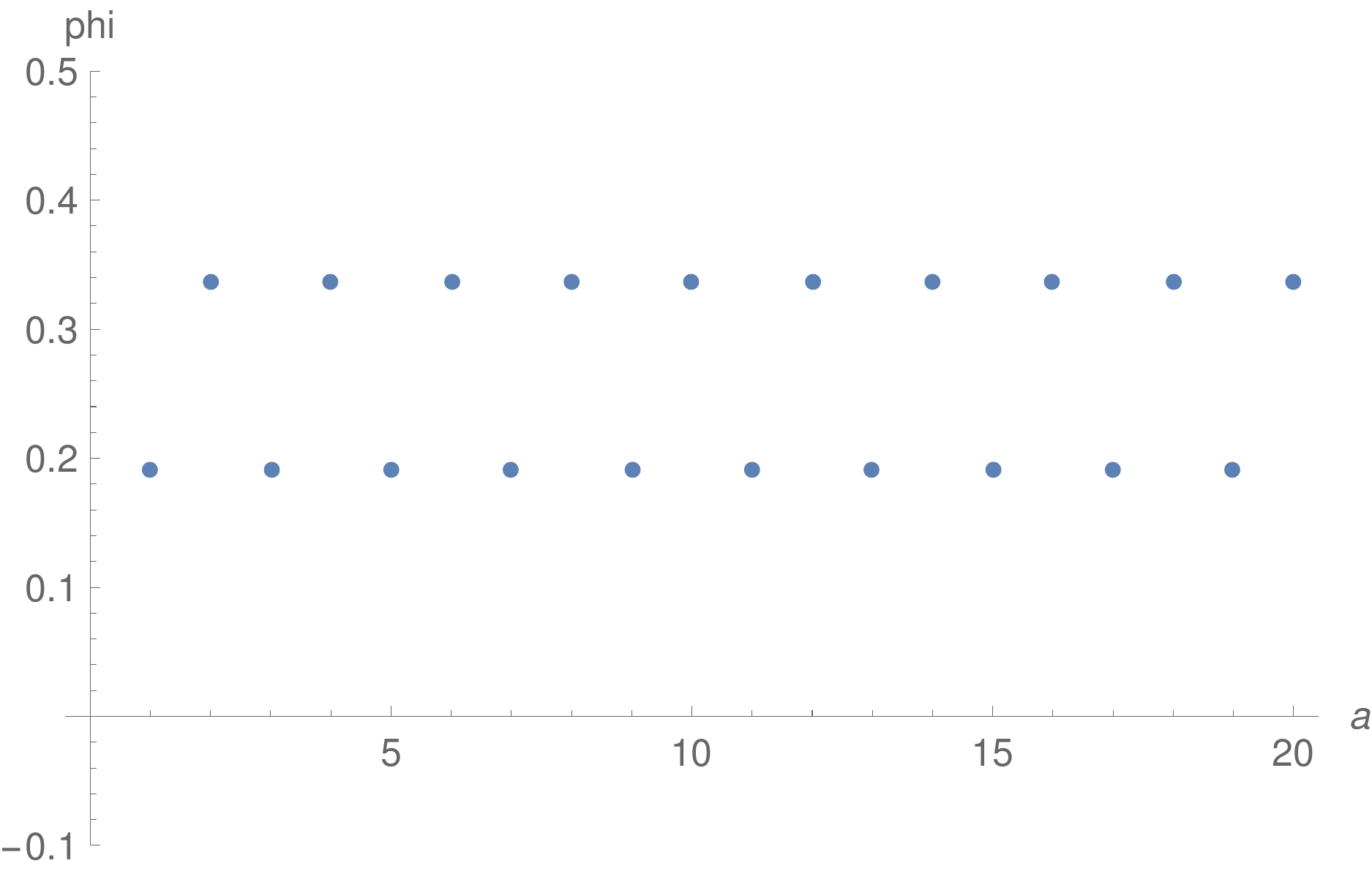}
\includegraphics[scale=.35]{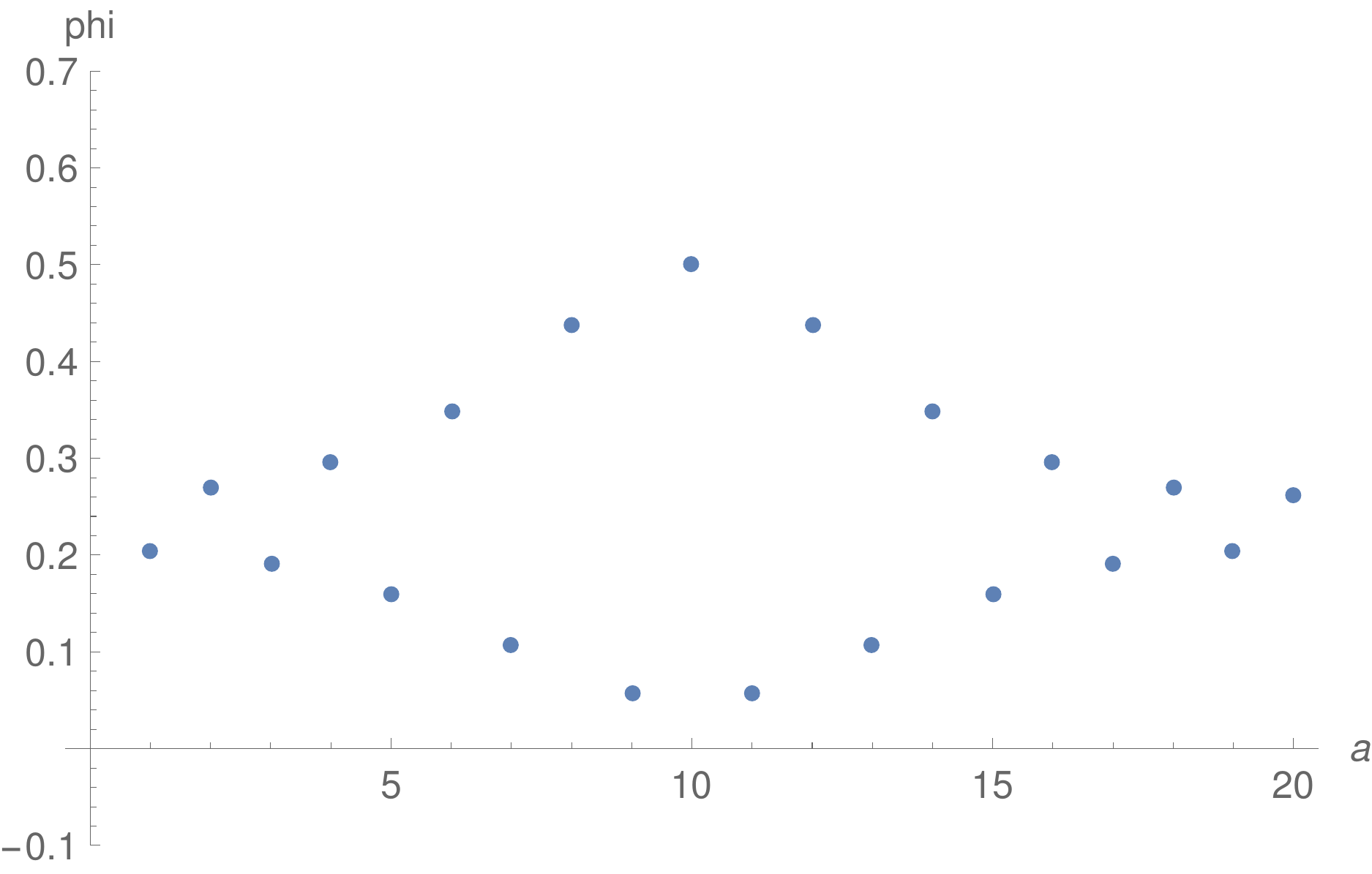}
\includegraphics[scale=.35]{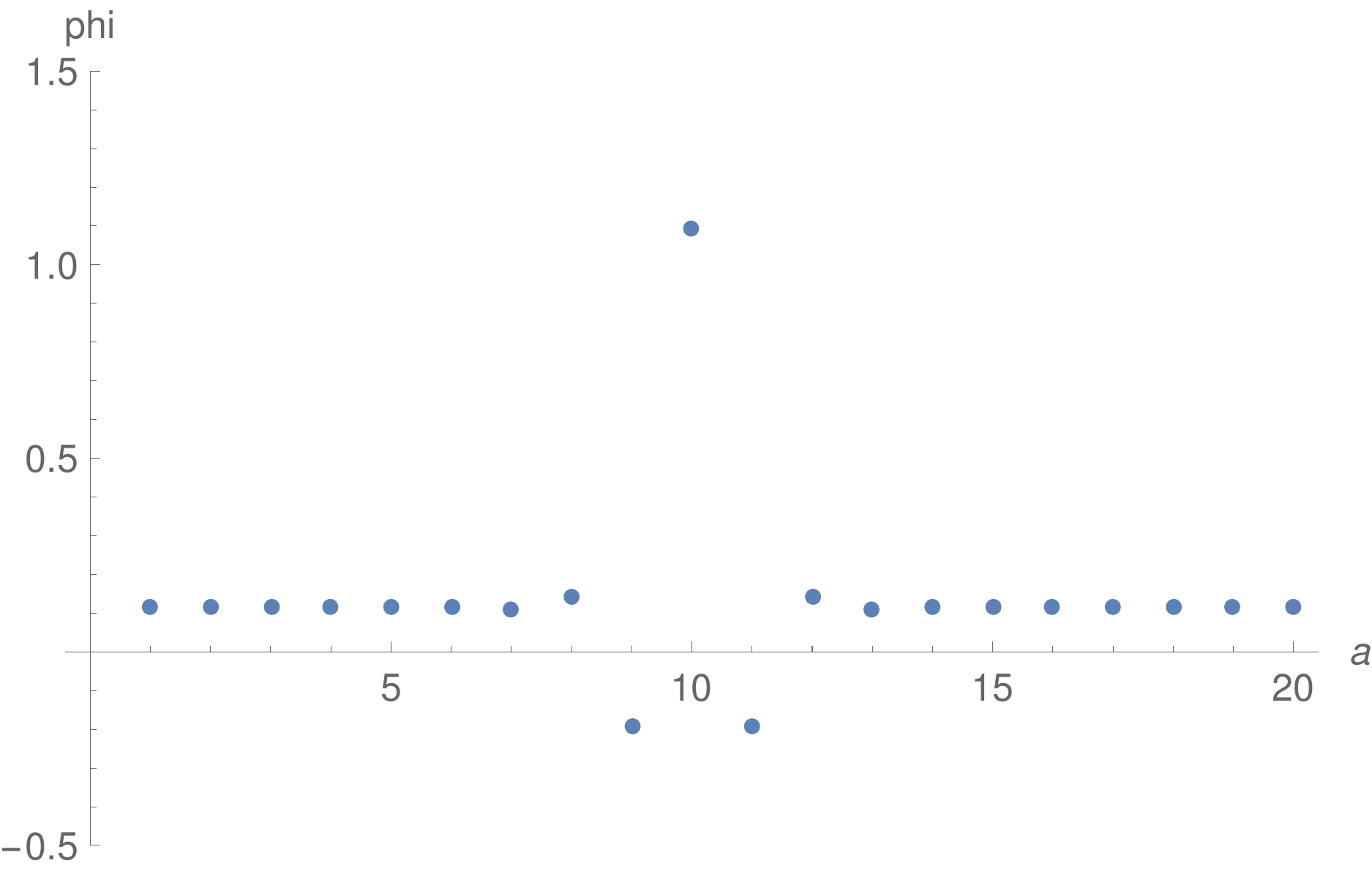}
\includegraphics[scale=.35]{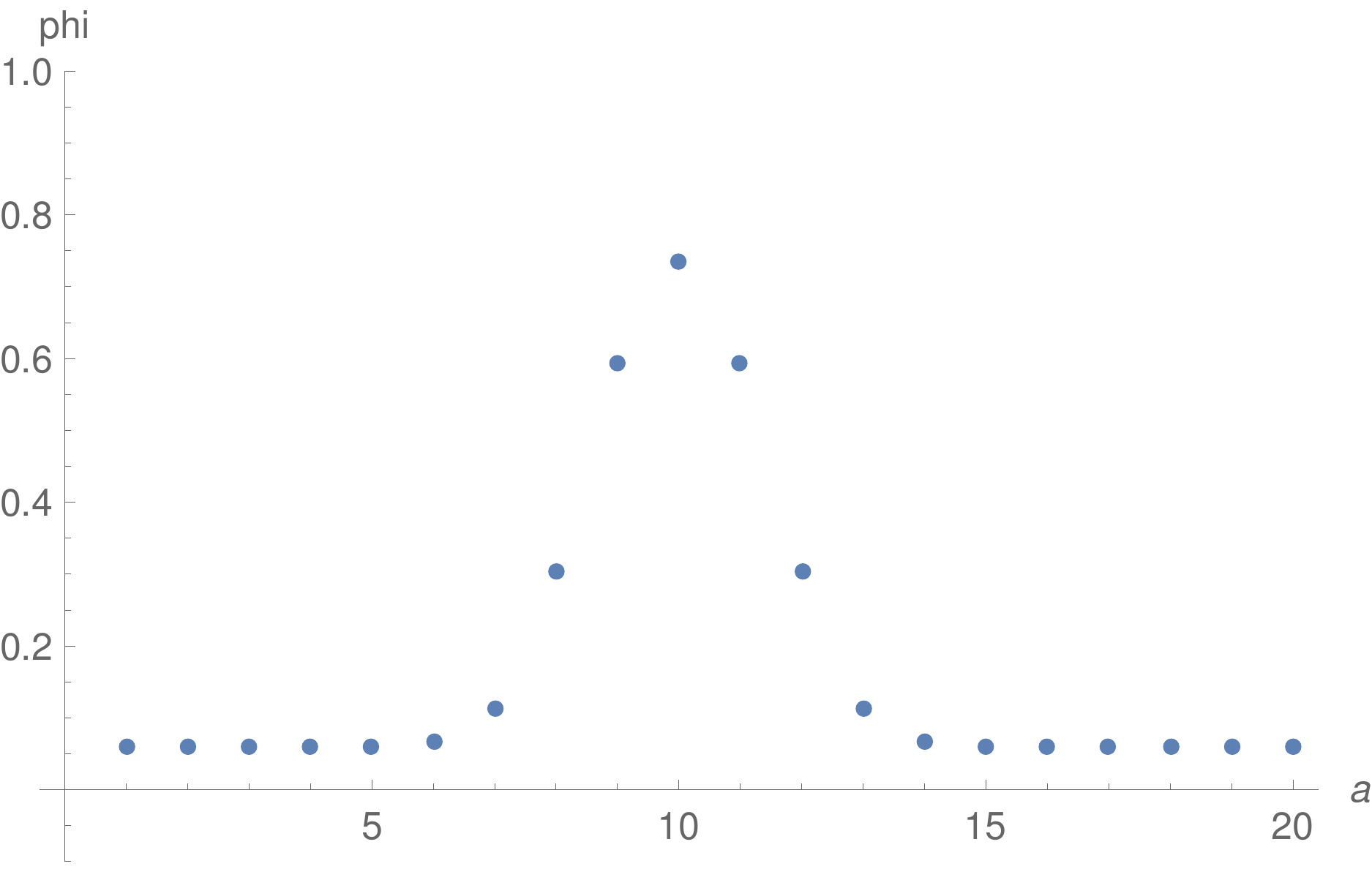}
\caption{The profiles of ${\bar \phi}$ for four values of $\xi$  with $N=20,\  \kappa=-1$. 
The up-left figure is for $\xi=-0.5$, the up-right for $\xi=-0.42$, 
the down-left for $\xi=-0.2$, and the down-right for $\xi=0.1$, respectively.
}
\label{fig:confphik=-1}
\end{center} 
\end{figure}

The phase depends on $N$ and $\kappa$ as well. 
The dependence does not seem simple, and some systematic studies 
seem to be required to obtain a convincing picture. 
Here, we just give another example for $N=20,\ \kappa=-1$ in Fig.~\ref{fig:freek=-1}. 
These figures respectively show the first and second derivatives of the free energy with respect to 
$\xi$, which are obtained by finite differentiations as in the previous example.   
There seem to exist more phases than the previous case with $\kappa=1$. 
In Fig.~\ref{fig:confphik=-1}, the profiles of ${\bar \phi}$ for four values of $\xi$ are
shown (The flat cases are not shown, since they are just constant). 
The down-right figure corresponds to the point-like space. The $\bar \phi$ in the first two figures 
are consistent with the picture that, as $\xi$ is increased from a large negative value,
two second order phase transitions triggered by negative eigenvalues 
occur successively.
The first one is a condensation of a mode with an oscillation period 2, and 
the second one of another mode with another frequency. 

\section{Summary and discussions}
\label{sec:summary}
In this paper, we have shown that 
classical spaces with geometries, generally curved, in arbitrary dimensions
can be generated on boundaries of randomly connected tensor networks
in the thermodynamic limit
by appropriately choosing the tensors. 
In particular, we have seen that 
such a tensor must contain not only a local part representing structures of 
local neighbourhoods but also a non-local part which stabilizes the space.
The non-local part does not cause problems of non-locality in the emergent space, 
since it affects only a single global mode and cannot be detected by a local observer
in the large $N$ limit.
We have given the explicit solvable examples of arbitrary dimensional flat tori. 
As for the general case, 
the correspondence from the tensor to the geometry of the emergent space
has been obtained from an analysis of  an effective action.  
The action has an invariant form under the spatial diffeomorphism, which is 
an outcome of the underlying orthogonal group symmetry of the randomly connected tensor network.
We have also studied phase transitions among various kinds of spaces
including extended and point-like ones.

In the randomly connected tensor network, the tensor is a given external variable
rather than a dynamical one, and therefore the emergent space is a static object. 
This is an unsatisfactory situation from the perspective of quantum gravity
and even in classical contexts.
On the other hand, the tensor is a dynamical variable 
in the canonical tensor model (CTM), and its Hamiltonian generates 
a flow equation for the tensor of the randomly connected tensor 
network, which can be regarded as a renormalization group flow \cite{Sasakura:2014zwa,Sasakura:2015xxa}. 
Then, by using the result of this paper about the correspondence from
the tensor to the geometry,  the dynamical equation of CTM can be translated 
to a dynamical equation for the geometry of the emergent space. 
The form of the latter equation would be highly interesting to study, 
since CTM is known to have intimate connections to general relativity
\cite{Sasakura:2014gia,Sasakura:2015pxa}.
It would also be interesting to see whether
the tensors generating extended spaces are dynamically favoured or not,
since quantum gravity should provide some explanations for our actual spacetime.    

Interestingly, the non-local part of the tensor has precisely
the form which is equivalent to an addition of a negative cosmological constant 
in the framework of CTM \cite{Narain:2014cya}. 
Then the two main implications of this paper,  
emergent spaces on boundaries and necessity of negative cosmological constants for stability,  
curiously resembles a part of the 
AdS/CFT correspondence \cite{Maldacena:1997re} in string theory.
It has also been argued that the tensor networks (though not random) are discrete 
realizations of the AdS spaces \cite{Swingle:2009bg}. 
Showing any connections between our framework and the AdS/CFT correspondence is 
presently far beyond our scope, but the resemblance would 
at least suggest an interesting direction of study:
not only boundaries, but random networks themselves may also have effective geometries, 
as the bulk geometries of the AdS spaces.  
This contradicts the difficulties mentioned in Section~\ref{sec:space},
but there would remain the possibility that
such classical geometries on randomly connected tensor networks would appear in 
certain sophisticated limits of the parameters, as the bulk geometries of AdS in
string theory
have definite meanings only in semi-classical limits.

Obviously, putting matters on the emergent space would also be an interesting direction
of study. 
This would be possible by considering tensors more complex than this paper, and
the super-extension discussed in \cite{Narain:2015owa}.

\vspace{1cm}
\centerline{\bf Acknowledgements} 
The work of N.S. is supported in part by JSPS KAKENHI Grant Number 15K05050. 
The work of Y.S. is funded under CUniverse research promotion project 
by Chulalongkorn University (grant reference CUAASC).

\appendix 
\section{Notes on covariance}
\label{sec:covariance}
As mentioned in Section \ref{sec:general}, in this appendix, 
we will show the linear relation between the moments and the covariant moments 
defined by (\ref{eq:momentexp}) and (\ref{eq:momentexpcovariant}), respectively.
This ensures that the two ways of description are equivalent within the order we are 
considering.
In order to show this, 
note that the test functions, $f_1$, $f_2$ and $f_3$, 
defined in (\ref{eq:momentexp}) and (\ref{eq:momentexpcovariant}) 
should transform as scalar half-densities under the spatial diffeomorphism \cite{Sasakura:2015pxa}. 
Thus we can parametrise the test functions in terms of scalars, $s_1$, $s_2$ and $s_3$, as
\[
f_1 = g^{1/4} s_1, \ \ \ 
f_2 = g^{1/4} s_2, \ \ \ 
f_3 = g^{1/4} s_3.
\label{eq:modifiedtestfunctions}
\] 
Plugging (\ref{eq:modifiedtestfunctions}) into (\ref{eq:momentexp}) and (\ref{eq:momentexpcovariant}) 
and comparing the coefficients of $s_1s_2s_3$, $s_1\partial_{\mu}(s_2s_3)$, $s_1\partial_{\mu}s_2 \partial_{\nu} s_3$, and
$s_1(\partial_{\mu}\partial_\nu s_2)s_3 + s_1 s_2 (\partial_\mu \partial_\nu s_3)$,
respectively,
we can obtain the relation between the moments and the covariant moments as
\[
&\alpha_c 
= \alpha + \frac{1}{2} \beta^{\mu}g_{\mu} 
+ \frac{1}{4} \gamma^{\mu \nu} \left( \partial_{\mu}g_{\nu} + \frac{1}{4} g_{\mu} g_{\nu}  \right) 
+ \frac{1}{16} \tilde \gamma^{\mu \nu}g_{\mu}g_{\nu} + \cdots, \\ 
&\beta^{\mu}_c-\frac{1}{2} \gamma_c^{\nu\rho} \Gamma_{\nu\rho}^\mu
= \beta^{\mu} 
+ \frac{1}{4} \gamma^{\mu \nu}g_{\nu} 
+ \frac{1}{4} \tilde \gamma^{\mu \nu} g_{\nu} 
+ \cdots, \\
&\gamma^{\mu \nu}_c=\gamma^{\mu \nu} +\cdots, \\
&\tilde \gamma^{\mu \nu}_c=\tilde \gamma^{\mu \nu}+\cdots,
\label{eq:relationmoments} 
\] 
where $\Gamma^{\mu}_{\nu \rho}$ is the Christoffel symbol associated with $g_{\mu \nu}$, 
the dots represent the higher order corrections neglected in the analysis and 
\[
g_{\mu} := g^{\lambda \tau}\partial_{\mu} g_{\lambda \tau}. 
\label{eq:gmu}
\]


\end{document}